\documentclass[apjl]{emulateapj}
\usepackage{comment}
\usepackage{ifthen}

\newcommand{\ctbd}[1]{}

\newcommand{\lc}{light curve}
\newcommand{\lcs}{light curves}
\newcommand{\Lc}{Light curve}

\newcommand{\band}[1]{\ensuremath{#1}-band}

\newcommand{\kms}{\ensuremath{\rm km\,s^{-1}}}
\newcommand{\ms}{\ensuremath{\rm m\,s^{-1}}}

\newcommand{\gcmc}{\ensuremath{\rm g\,cm^{-3}}}
\newcommand{\ergscmsq}{\ensuremath{\rm erg\,s^{-1}\,cm^{-2}}}

\newcommand{\vsini}{\ensuremath{v \sin{i}}}
\newcommand{\feh}{\ensuremath{\rm [Fe/H]}}

\newcommand{\vmac}{\ensuremath{v_{\rm mac}}}
\newcommand{\vmic}{\ensuremath{v_{\rm mic}}}
\newcommand{\rhk}{\ensuremath{R^{\prime}_{HK}}}
\newcommand{\logrhk}{\ensuremath{\log\rhk}}

\newcommand{\rsun}{\ensuremath{R_\sun}}
\newcommand{\msun}{\ensuremath{M_\sun}}
\newcommand{\lsun}{\ensuremath{L_\sun}}

\newcommand{\rstar}{\ensuremath{R_\star}}
\newcommand{\mstar}{\ensuremath{M_\star}}
\newcommand{\lstar}{\ensuremath{L_\star}}

\newcommand{\teffstar}{\ensuremath{T_{\rm eff\star}}}

\newcommand{\loggstar}{\ensuremath{\log{g_{\star}}}}

\newcommand{\mearth}{\ensuremath{M_\earth}}

\newcommand{\rpl}{\ensuremath{R_{p}}}
\newcommand{\mpl}{\ensuremath{M_{p}}}

\newcommand{\rhopl}{\ensuremath{\rho_{p}}}

\newcommand{\arstar}{\ensuremath{a/\rstar}}
\newcommand{\zrstar}{\ensuremath{\zeta/\rstar}}

\newcommand{\rjup}{\ensuremath{R_{\rm J}}}
\newcommand{\mjup}{\ensuremath{M_{\rm J}}}

\newcommand{\refsec}[1]{\mbox{\S\ \ref{sec:#1}}}

\newcommand{\refsecl}[1]{\mbox{Section \ref{sec:#1}}}

\newcommand{\reftabl}[1]{Table~\ref{tab:#1}}

\newcommand{\flwof}{\mbox{FLWO 1.2\,m}}

\newcommand{\flwos}{\mbox{FLWO 1.5\,m}}

\newcommand{\hatcurhtr}{HTR164-005}                                    
\newcommand{\hatcurfieldA}{163}                                         
\newcommand{\hatcurfieldB}{164}                                         
\newcommand{\hatcurCCra}{\ensuremath{00^{\mathrm h}52^{\mathrm m}00.18{\mathrm s}}}                                  
\newcommand{\hatcurCCdec}{\ensuremath{+34{\arcdeg}43{\arcmin}42.3{\arcsec}}}                                 
\newcommand{\hatcurCCtwomass}{2MASS~00520018+3443422}                  
\newcommand{\hatcurCCgsc}{GSC~2284-00503}                              
\newcommand{\hatcurCCtassmv}{13.03}                                   
\newcommand{\hatcurCCtwomassJmag}{\ensuremath{11.561\pm0.020}}         
\newcommand{\hatcurCCtwomassHmag}{\ensuremath{11.206\pm0.023}}         
\newcommand{\hatcurCCtwomassKmag}{\ensuremath{11.104\pm0.018}}         
\newcommand{\hatcurCCesoJKmag}{\ensuremath{0.487\pm0.030}}             
\newcommand{\hatcurLCdip}{\ensuremath{16.1}}                           
\newcommand{\hatcurLCrprstar}{\ensuremath{0.1130\pm0.0024}}            
\newcommand{\hatcurLCbsq}{\ensuremath{0.089_{-0.052}^{+0.087}}}        
\newcommand{\hatcurLCimp}{\ensuremath{0.299_{-0.123}^{+0.114}}}        
\newcommand{\hatcurLCzeta}{\ensuremath{16.77\pm0.16}}                  
\newcommand{\hatcurLCdur}{\ensuremath{0.1341\pm0.0020}}                
\newcommand{\hatcurLCingdur}{\ensuremath{0.0148\pm0.0015}}             
\newcommand{\hatcurLCP}{\ensuremath{3.257215\pm0.000007}}              
\newcommand{\hatcurLCPshort}{\ensuremath{3.2572}}                      
\newcommand{\hatcurLCT}{\ensuremath{2455417.59832\pm0.00053}}          
\newcommand{\hatcurSMEiteff}{\ensuremath{5680\pm90}}                   
\newcommand{\hatcurSMEizfeh}{\ensuremath{+0.12\pm0.08}}                 
\newcommand{\hatcurSMEizfehshort}{\ensuremath{+0.12}}                   
\newcommand{\hatcurSMEilogg}{\ensuremath{4.42\pm0.06}}                 
\newcommand{\hatcurSMEivsin}{\ensuremath{0.2_{-0.2}^{+0.5}}}                   
\newcommand{\hatcurSMEivmac}{\ensuremath{3.88}}                        
\newcommand{\hatcurSMEivmic}{\ensuremath{0.85}}                        
\newcommand{\hatcurSMEiiteff}{\ensuremath{5681\pm88}}                  
\newcommand{\hatcurSMEiizfeh}{\ensuremath{+0.12\pm0.08}}                
\newcommand{\hatcurSMEiizfehshort}{\ensuremath{+0.12}}                  
\newcommand{\hatcurSMEiilogg}{\ensuremath{4.42\pm0.06}}                
\newcommand{\hatcurSMEiivsin}{\ensuremath{0.2\pm0.5}}                  
\newcommand{\hatcurSMEiivmac}{\ensuremath{NULL}}                       
\newcommand{\hatcurSMEiivmic}{\ensuremath{NULL}}                       
\newcommand{\hatcurFIESgamma}{\ensuremath{+43.884\pm0.068}}             
\newcommand{\hatcurLBii}{\ensuremath{0.2879}}                          
\newcommand{\hatcurLBiii}{\ensuremath{0.3220}}                         
\newcommand{\hatcurISOmlong}{\ensuremath{1.025\pm0.047}}               
\newcommand{\hatcurISOrlong}{\ensuremath{1.103_{-0.069}^{+0.091}}}     
\newcommand{\hatcurISOlogg}{\ensuremath{4.36\pm0.06}}                  
\newcommand{\hatcurISOlum}{\ensuremath{1.13_{-0.16}^{+0.23}}}          
\newcommand{\hatcurISOmv}{\ensuremath{4.70\pm0.19}}                    
\newcommand{\hatcurISOage}{\ensuremath{6.1_{-1.9}^{+2.6}}}             
\newcommand{\hatcurISOMK}{\ensuremath{3.11\pm0.16}}                    
\newcommand{\hatcurISOJK}{\ensuremath{0.410\pm0.020}}                    
\newcommand{\hatcurISOspec}{G3}                                        
\newcommand{\hatcurRVK}{\ensuremath{84.7\pm4.2}}                       
\newcommand{\hatcurRVk}{\ensuremath{-0.010\pm0.028}}                   
\newcommand{\hatcurRVh}{\ensuremath{-0.022\pm0.053}}                   
\newcommand{\hatcurRVjitterA}{\ensuremath{15.6}}                       
\newcommand{\hatcurRVfitrmsA}{\ensuremath{16.1}}                       
\newcommand{\hatcurRVfitrmsB}{\ensuremath{10.1}}                       
\newcommand{\hatcurRVeccen}{\ensuremath{0.051\pm0.033}}                
\newcommand{\hatcurRVomega}{\ensuremath{233\pm90}}                     
\newcommand{\hatcurPPi}{\ensuremath{88.0\pm0.9}}                       
\newcommand{\hatcurPPlogg}{\ensuremath{3.02\pm0.06}}                   
\newcommand{\hatcurPPar}{\ensuremath{8.43\pm0.57}}                     
\newcommand{\hatcurPParel}{\ensuremath{0.0434\pm0.0007}}               
\newcommand{\hatcurPPrho}{\ensuremath{0.44\pm0.09}}                    
\newcommand{\hatcurPPm}{\ensuremath{0.63\pm0.04}}                      
\newcommand{\hatcurPPmlong}{\ensuremath{0.626\pm0.037}}                
\newcommand{\hatcurPPr}{\ensuremath{1.21_{-0.08}^{+0.11}}}             
\newcommand{\hatcurPPrlong}{\ensuremath{1.212_{-0.082}^{+0.113}}}      
\newcommand{\hatcurPPmrcorr}{\ensuremath{0.46}}                        
\newcommand{\hatcurPPteff}{\ensuremath{1384\pm52}}                     
\newcommand{\hatcurPPtheta}{\ensuremath{0.044\pm0.003}}                
\newcommand{\hatcurPPfluxperi}{\ensuremath{9.16_{-0.92}^{+1.80}}}      
\newcommand{\hatcurPPfluxperidim}{\ensuremath{8}}                      
\newcommand{\hatcurPPfluxap}{\ensuremath{7.58\pm1.34}}                 
\newcommand{\hatcurPPfluxapdim}{\ensuremath{8}}                        
\newcommand{\hatcurPPfluxavg}{\ensuremath{8.28_{-1.06}^{+1.49}}}       
\newcommand{\hatcurPPfluxavgdim}{\ensuremath{8}}                       
\newcommand{\hatcurXsecondary}{\ensuremath{2455419.205\pm0.058}}       
\newcommand{\hatcurXsecdur}{\ensuremath{0.1289\pm0.0126}}              
\newcommand{\hatcurXsecingdur}{\ensuremath{0.0142\pm0.0023}}           
\newcommand{\hatcurXdist}{\ensuremath{395_{-26}^{+34}}}                

\newcommand{\hatcur}{HAT-P-28}
\newcommand{\hatcurb}{HAT-P-28b}

\newcommand{\hatcurCCtassvi}{\ensuremath{0.85\pm0.18}}                  
\newcommand{\hatcurSMEversion}{i}                                       
\newcommand{\hatcurSMEteff}{\ifthenelse{\equal{\hatcurSMEversion}{i}}{\hatcurSMEiteff}{\hatcurSMEiiteff}}
\newcommand{\hatcurSMEzfeh}{\ifthenelse{\equal{\hatcurSMEversion}{i}}{\hatcurSMEizfeh}{\hatcurSMEiizfeh}}
\newcommand{\hatcurSMEzfehshort}{\ifthenelse{\equal{\hatcurSMEversion}{i}}{\hatcurSMEizfehshort}{\hatcurSMEiizfehshort}}
\newcommand{\hatcurSMElogg}{\ifthenelse{\equal{\hatcurSMEversion}{i}}{\hatcurSMEilogg}{\hatcurSMEiilogg}}
\newcommand{\hatcurSMEvsin}{\ifthenelse{\equal{\hatcurSMEversion}{i}}{\hatcurSMEivsin}{\hatcurSMEiivsin}}
\newcommand{\hatcurSMEvmac}{\ifthenelse{\equal{\hatcurSMEversion}{i}}{\hatcurSMEivmac}{\hatcurSMEiivmac}}
\newcommand{\hatcurSMEvmic}{\ifthenelse{\equal{\hatcurSMEversion}{i}}{\hatcurSMEivmic}{\hatcurSMEiivmic}}

\newcommand{\hatcurBBhtr}{HTR089-011}                                    
\newcommand{\hatcurBBfield}{089}                                         
\newcommand{\hatcurBBCCra}{\ensuremath{02^{\mathrm h}12^{\mathrm m}31.46{\mathrm s}}}                                  
\newcommand{\hatcurBBCCdec}{\ensuremath{+51{\arcdeg}46{\arcmin}43.5{\arcsec}}}                                 
\newcommand{\hatcurBBCCtwomass}{2MASS~02123147+5146435}                  
\newcommand{\hatcurBBCCgsc}{GSC~3293-01539}                              
\newcommand{\hatcurBBCCtassmv}{11.90}                                    
\newcommand{\hatcurBBCCtwomassJmag}{\ensuremath{10.648\pm0.023}}         
\newcommand{\hatcurBBCCtwomassHmag}{\ensuremath{10.395\pm0.023}}         
\newcommand{\hatcurBBCCtwomassKmag}{\ensuremath{10.297\pm0.020}}         
\newcommand{\hatcurBBCCesoJKmag}{\ensuremath{0.375\pm0.033}}             
\newcommand{\hatcurBBLCdip}{\ensuremath{9.1}}                            
\newcommand{\hatcurBBLCrprstar}{\ensuremath{0.0927\pm0.0028}}            
\newcommand{\hatcurBBLCbsq}{\ensuremath{0.349_{-0.095}^{+0.079}}}        
\newcommand{\hatcurBBLCimp}{\ensuremath{0.591_{-0.094}^{+0.062}}}        
\newcommand{\hatcurBBLCzeta}{\ensuremath{16.22\pm0.93}}                  
\newcommand{\hatcurBBLCdur}{\ensuremath{0.1407\pm0.0074}}                
\newcommand{\hatcurBBLCingdur}{\ensuremath{0.0177\pm0.0024}}             
\newcommand{\hatcurBBLCP}{\ensuremath{5.723186\pm0.000049}}              
\newcommand{\hatcurBBLCPshort}{\ensuremath{5.7232}}                      
\newcommand{\hatcurBBLCT}{\ensuremath{2455197.57540\pm0.00181}}          
\newcommand{\hatcurBBSMEiteff}{\ensuremath{6087\pm88}}                   
\newcommand{\hatcurBBSMEizfeh}{\ensuremath{0.21\pm0.08}}                 
\newcommand{\hatcurBBSMEizfehshort}{\ensuremath{0.21}}                   
\newcommand{\hatcurBBSMEilogg}{\ensuremath{4.31\pm0.06}}                 
\newcommand{\hatcurBBSMEivsin}{\ensuremath{3.9\pm0.5}}                   
\newcommand{\hatcurBBSMEivmac}{\ensuremath{4.50}}                        
\newcommand{\hatcurBBSMEivmic}{\ensuremath{0.85}}                        
\newcommand{\hatcurBBSMEiiteff}{\ensuremath{6087\pm88}}                  
\newcommand{\hatcurBBSMEiizfeh}{\ensuremath{0.21\pm0.08}}                
\newcommand{\hatcurBBSMEiizfehshort}{\ensuremath{0.21}}                  
\newcommand{\hatcurBBSMEiilogg}{\ensuremath{4.31\pm0.06}}                
\newcommand{\hatcurBBSMEiivsin}{\ensuremath{3.9\pm0.5}}                  
\newcommand{\hatcurBBSMEiivmac}{\ensuremath{4.50}}                       
\newcommand{\hatcurBBSMEiivmic}{\ensuremath{0.85}}                       
\newcommand{\hatcurBBFIESgamma}{\ensuremath{-21.670\pm0.08}}             
\newcommand{\hatcurBBLBii}{\ensuremath{0.2273}}                          
\newcommand{\hatcurBBLBiii}{\ensuremath{0.3581}}                         
\newcommand{\hatcurBBISOmlong}{\ensuremath{1.207\pm0.046}}               
\newcommand{\hatcurBBISOrlong}{\ensuremath{1.224_{-0.075}^{+0.133}}}     
\newcommand{\hatcurBBISOlogg}{\ensuremath{4.34\pm0.06}}                  
\newcommand{\hatcurBBISOlum}{\ensuremath{1.84_{-0.26}^{+0.47}}}          
\newcommand{\hatcurBBISOmv}{\ensuremath{4.11\pm0.21}}                    
\newcommand{\hatcurBBISOage}{\ensuremath{2.2\pm1.0}}                     
\newcommand{\hatcurBBISOMK}{\ensuremath{2.77\pm0.19}}                    
\newcommand{\hatcurBBISOJK}{\ensuremath{0.340\pm0.020}}                    
\newcommand{\hatcurBBISOspec}{F8}                                         
\newcommand{\hatcurBBRVK}{\ensuremath{78.3\pm5.9}}                       
\newcommand{\hatcurBBRVk}{\ensuremath{-0.084_{-0.046}^{+0.026}}}         
\newcommand{\hatcurBBRVh}{\ensuremath{0.016\pm0.058}}                    
\newcommand{\hatcurBBRVjitter}{\ensuremath{6.0}}                         
\newcommand{\hatcurBBRVfitrms}{\ensuremath{6.5}}                         
\newcommand{\hatcurBBRVeccen}{\ensuremath{0.095\pm0.047}}                
\newcommand{\hatcurBBRVomega}{\ensuremath{169\pm30}}                     
\newcommand{\hatcurBBPPi}{\ensuremath{87.1_{-0.7}^{+0.5}}}               
\newcommand{\hatcurBBPPlogg}{\ensuremath{3.20\pm0.07}}                   
\newcommand{\hatcurBBPPar}{\ensuremath{11.70_{-0.97}^{+0.71}}}           
\newcommand{\hatcurBBPParel}{\ensuremath{0.0667\pm0.0008}}               
\newcommand{\hatcurBBPPrho}{\ensuremath{0.71\pm0.18}}                    
\newcommand{\hatcurBBPPm}{\ensuremath{0.78_{-0.04}^{+0.08}}}             
\newcommand{\hatcurBBPPmlong}{\ensuremath{0.778_{-0.040}^{+0.076}}}      
\newcommand{\hatcurBBPPr}{\ensuremath{1.11_{-0.08}^{+0.14}}}             
\newcommand{\hatcurBBPPrlong}{\ensuremath{1.107_{-0.082}^{+0.136}}}      
\newcommand{\hatcurBBPPmrcorr}{\ensuremath{0.54}}                        
\newcommand{\hatcurBBPPteff}{\ensuremath{1260_{-45}^{+64}}}              
\newcommand{\hatcurBBPPtheta}{\ensuremath{0.077\pm0.007}}                
\newcommand{\hatcurBBPPfluxperi}{\ensuremath{6.90_{-1.01}^{+3.61}}}      
\newcommand{\hatcurBBPPfluxap}{\ensuremath{4.72_{-0.62}^{+0.81}}}        
\newcommand{\hatcurBBPPfluxavg}{\ensuremath{5.69_{-0.75}^{+1.36}}}       
\newcommand{\hatcurBBXsecondary}{\ensuremath{2455200.132\pm0.138}}       
\newcommand{\hatcurBBXsecdur}{\ensuremath{0.1424\pm0.0107}}              
\newcommand{\hatcurBBXsecingdur}{\ensuremath{0.0183\pm0.0074}}           
\newcommand{\hatcurBBXdist}{\ensuremath{322_{-21}^{+35}}}                

\newcommand{\hatcurBB}{HAT-P-29}
\newcommand{\hatcurBBb}{HAT-P-29b}

\newcommand{\hatcurBBCCtassvi}{\ensuremath{0.85\pm0.18}}                  
\newcommand{\hatcurBBSMEversion}{i}                                       
\newcommand{\hatcurBBSMEteff}{\ifthenelse{\equal{\hatcurBBSMEversion}{i}}{\hatcurBBSMEiteff}{\hatcurBBSMEiiteff}}
\newcommand{\hatcurBBSMEzfeh}{\ifthenelse{\equal{\hatcurBBSMEversion}{i}}{\hatcurBBSMEizfeh}{\hatcurBBSMEiizfeh}}
\newcommand{\hatcurBBSMEzfehshort}{\ifthenelse{\equal{\hatcurBBSMEversion}{i}}{\hatcurBBSMEizfehshort}{\hatcurBBSMEiizfehshort}}
\newcommand{\hatcurBBSMElogg}{\ifthenelse{\equal{\hatcurBBSMEversion}{i}}{\hatcurBBSMEilogg}{\hatcurBBSMEiilogg}}
\newcommand{\hatcurBBSMEvsin}{\ifthenelse{\equal{\hatcurBBSMEversion}{i}}{\hatcurBBSMEivsin}{\hatcurBBSMEiivsin}}
\newcommand{\hatcurBBSMEvmac}{\ifthenelse{\equal{\hatcurBBSMEversion}{i}}{\hatcurBBSMEivmac}{\hatcurBBSMEiivmac}}
\newcommand{\hatcurBBSMEvmic}{\ifthenelse{\equal{\hatcurBBSMEversion}{i}}{\hatcurBBSMEivmic}{\hatcurBBSMEiivmic}}

\newcommand{\hatcurisoshort}{YY}
\newcommand{\hatcurisofull}{Yonsei-Yale (YY)}
\newcommand{\hatcurisocite}{yi:2001}
\newcommand{\hatcurlumind}{\arstar}
\newcommand{\hatcurjhkfilset}{ESO}

\newboolean{emulateapj}
\setboolean{emulateapj}{true}

\newboolean{rvtablelong}
\setboolean{rvtablelong}{true}

\newboolean{astroph}
\setboolean{astroph}{true}

\shortauthors{Buchhave et al.}
\shorttitle{\hatcur\lowercase{b} and \hatcurBB\lowercase{b}}
\ifthenelse{\boolean{emulateapj}}{
    \newcommand{\titledag}{$\dagger$}
}{
    \newcommand{\titledag}{\dagger}
}

\begin{document}

\title{\hatcur\lowercase{b} and \hatcurBB\lowercase{b}: Two sub-Jupiter mass transiting planets
	\altaffilmark{$\star$},\altaffilmark{\titledag}}

\author{
   L.~A.~Buchhave\altaffilmark{1,2},
   G.~\'A.~Bakos\altaffilmark{1,3},
   J.~D.~Hartman\altaffilmark{1},
   G.~Torres\altaffilmark{1},
   D.~W.~Latham\altaffilmark{1},
   J.~Andersen\altaffilmark{2,9},
   G.~Kov\'acs\altaffilmark{4},
   R.~W.~Noyes\altaffilmark{1},
   A.~Shporer\altaffilmark{10,11},
   G.~A.~Esquerdo\altaffilmark{1},
   D.~A.~Fischer\altaffilmark{5},
   J.~A.~Johnson\altaffilmark{6},
   G.~W.~Marcy\altaffilmark{7},
   A.~W.~Howard\altaffilmark{7},
   B.~B\'eky\altaffilmark{1} 
   D.~D.~Sasselov\altaffilmark{1},
   G.~F\H{u}r\'esz\altaffilmark{1},
   S.~N.~Quinn\altaffilmark{1},
   R.~P.~Stefanik\altaffilmark{1},
   T.~Szklen\'ar\altaffilmark{1},
   P.~Berlind\altaffilmark{1},
   M.~L.~Calkins\altaffilmark{1},
   J.~L\'az\'ar\altaffilmark{8},
   I.~Papp\altaffilmark{8},
   P.~S\'ari\altaffilmark{8}
}

\altaffiltext{1}{Harvard-Smithsonian Center for Astrophysics, 
	Cambridge, MA}

\altaffiltext{2}{Niels Bohr Institute, Copenhagen University, Denmark}

\altaffiltext{3}{NSF Fellow}

\altaffiltext{4}{Konkoly Observatory, Budapest, Hungary}

\altaffiltext{5}{Department of Astronomy, Yale University, New Haven, CT}

\altaffiltext{6}{Department of Astrophysics, California Institute of 
Technology, Pasadena, CA, USA}

\altaffiltext{7}{Department of Astronomy, University of California,
	Berkeley, CA}

\altaffiltext{8}{Hungarian Astronomical Association, Budapest, 
	Hungary}

\altaffiltext{9}{Nordic Optical Telescope Scientific Association, 
La Palma, Canarias, Spain}

\altaffiltext{10}{Las Cumbres Observatory Global Telescope Network, Goleta, CA, USA}

\altaffiltext{11}{Department of Physics, University of California, Santa Barbara, CA, USA}

\altaffiltext{$\star$}{
	Based in part on observations made with the Nordic Optical
        Telescope, operated on the island of La Palma jointly by
        Denmark, Finland, Iceland, Norway, and Sweden, in the Spanish
        Observatorio del Roque de los Muchachos of the Instituto de
        Astrofisica de Canarias.
}
\altaffiltext{$\dagger$}{
	Based in part on observations obtained at the W.~M.~Keck
	Observatory, which is operated by the University of California and
	the California Institute of Technology. Keck time has been
	granted by NOAO (A201Hr) and NASA (N018Hr, N167Hr).
}


\begin{abstract}
\setcounter{footnote}{10}

We present the discovery of two transiting exoplanets. \hatcurb{} orbits a V=\hatcurCCtassmv\ \hatcurISOspec\ dwarf star with a period $P=\hatcurLCPshort$\,d and has a mass of \hatcurPPm\,\mjup\ and a radius of \hatcurPPr\,\rjup\ yielding a mean density of \hatcurPPrho\ \gcmc. \hatcurBBb{} orbits a V=\hatcurBBCCtassmv\ \hatcurBBISOspec\ dwarf star with a period $P=\hatcurBBLCPshort$\,d and has a mass of \hatcurBBPPm\,\mjup\ and a radius of \hatcurBBPPr\,\rjup\ yielding a mean density of \hatcurBBPPrho\ \gcmc. We discuss the properties of these planets in the
context of other known transiting planets.
\setcounter{footnote}{0}
\end{abstract}

\keywords{
	planetary systems ---
	stars: individual (\hatcur{}, \hatcurCCgsc{}, \hatcurBB{}, \hatcurBBCCgsc{}) 
	techniques: spectroscopic, photometric
}


\section{Introduction}
\label{sec:introduction}


When a planet transits its host star, we can detect not only the
presence of the planet, but also learn a great deal about its
physical properties: we can infer the planet's mass and radius and
thus its bulk density and composition, its temperature and atmospheric
composition \citep{charbonneau:2005}, and the dynamics of the
planetary system such as the projected spin-orbit alignment
\citep{winn:2005}. With over 100 transiting exoplanets
(TEPs) discovered in wide-angle transit surveys and the wealth of transiting planet candidates from the Kepler mission \citep{borucki:2011}, statistical analyzes are beginning to support constraints on
the planet formation models \citep{triaud:2010,morton:2011}.

Here we report the discovery of two new TEPs, \hatcurb\ and \hatcurBBb, orbiting the stars \hatcurCCgsc{} and \hatcurBBCCgsc{}, respectively. The Hungarian-made Automated Telescope Network \citep[HATNet;][]{bakos:2004} has been in operation since
2003, and has now covered approximately 14\% of the sky, searching
for TEPs around bright stars ($9.5\lesssim r \lesssim 14.5$). Although \hatcurb\ and \hatcurBBb\ are of the most common types of confirmed TEPs with respect to their mass and radii, they contribute to the characterization of the growing
population of Jupiter--sized planets known to transit.  The two discovered planets reported here orbit relatively faint stars compared to the current population of validated transiting planets, but the typical HATNet planets have relatively bright host stars compared to the wealth of planets candidates discovered by Kepler, and are hence more amenable to follow-up confirmation and characterization.

In \refsecl{obs} we briefly describe the
initial photometric detection and subsequent spectroscopic and
photometric follow-up observations used to confirm the planetary
nature of \hatcurb{} and \hatcurBBb{}. In \refsecl{analysis} we describe the data
analysis which we performed to determine the planetary and stellar
parameters. Our findings are discussed in \refsecl{discussion}.

\section{Observations}
\label{sec:obs}

\hatcurb{} and \hatcurBBb{} were discovered by HATNet following an observational
procedure which has been described in detail in several previous
discovery papers \citep[e.g.][]{bakos:2010,buchhave:2010}. Briefly, the
procedure consists of the following steps:
\begin{enumerate}
\item Photometric observations with one or more of the HATNet
  telescopes to detect stars with periodic transit-like light curves.
\item High-resolution, low-S/N ``reconnaissance'' spectroscopy
  observations to reject many false positives (such as M dwarf stars
  transiting G or F stars, or eclipsing binary stars blended with
  brighter giant stars).
\item Higher-precision photometric observations during transit to
  confirm the detection, show that it is consistent with a transiting
  planet, and use in determining the system parameters.
\item High-resolution, high-S/N ``confirmation'' spectroscopy to
  detect the orbital motion of the star due to the planet,
  characterize the host star, and rule-out subtle blend scenarios.
\end{enumerate}
Below we describe details of this procedure that are
pertinent specifically to the discovery of \hatcurb{} and \hatcurBBb{}.

\subsection{Photometric detection}
\label{sec:detection}

\begin{figure}[!ht]
\plotone{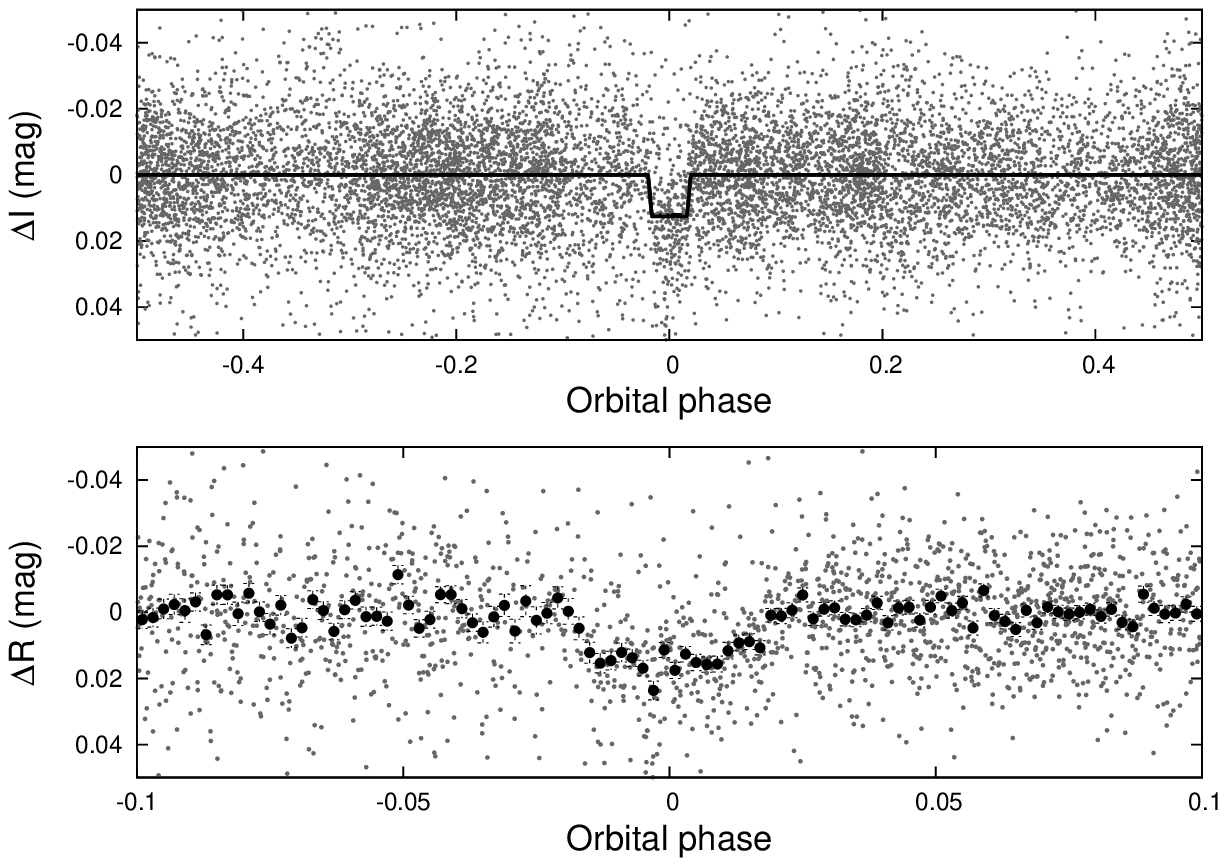}
\caption{
	Unbinned \lc{} of \hatcur{} including all 10,400 instrumental
        \band{R} 5.5 minute cadence measurements obtained with the
        HAT-6, HAT-7, HAT-8 and HAT-9 telescopes of HATNet (see
        \reftabl{photobs}), and folded with the period $P =
        \hatcurLCP$\,days resulting from the global fit described
        in \refsecl{analysis}.  The solid line shows a simplified
        transit model fit to the light curve (\refsecl{globmod}). The
        lower panel shows a zoomed-in view of the transit; the dark
        filled points show the light curve binned in phase using a
        bin-size of 0.002.
\label{fig:hatnet}}
\end{figure}
\begin{figure}[!ht]
\plotone{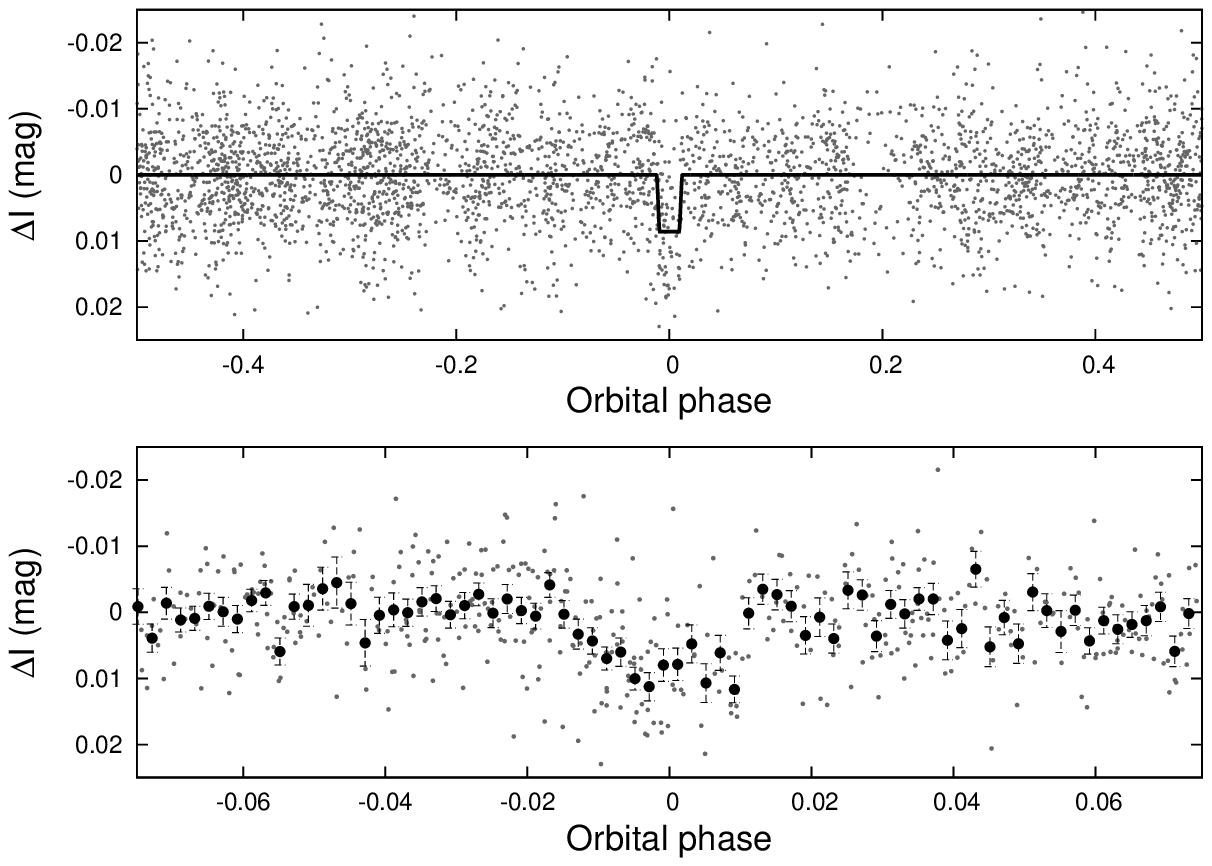}
\caption{
	Unbinned \lc{} of \hatcurBB{} including all 3,453 instrumental
        Sloan~$r$ 5.5 minute cadence measurements obtained with the
        HAT-5 and HAT-8 telescopes and folded with the period $P =
        \hatcurBBLCP$\,days. See Figure \ref{fig:hatnet} caption for details.
\label{fig:hatnetBB}}
\end{figure}

\ifthenelse{\boolean{emulateapj}}{
    \begin{deluxetable*}{lllrrr}
}{
    \begin{deluxetable}{lllrrr}
}
\tablewidth{0pc}
\tabletypesize{\scriptsize}
\tablecaption{
    Summary of photometric observations
    \label{tab:photobs}
}
\tablehead{
    \multicolumn{2}{c}{Instrument/Field\tablenotemark{a}}          &
    \multicolumn{1}{c}{Date(s)}             &
    \multicolumn{1}{c}{Number of Images}         &
    \multicolumn{1}{c}{Cadence (s)}         &
    \multicolumn{1}{c}{Filter}            \\
    &
    &
    &
    &
}
\startdata
\sidehead{\bf{\hatcurb}}
&HAT-7/163 & 2007 Sep--2008 Jan & 2329 & 330 & $R$ \\
&HAT-8/163 & 2007 Sep--2008 Jan & 1624 & 330 & $R$ \\
&HAT-6/164 & 2007 Sep--2008 Feb & 3685 & 330 & $R$ \\
&HAT-9/164 & 2007 Sep--2008 Feb & 2779 & 330 & $R$ \\
&KeplerCam\tablenotemark{b} & 2010 Sep 03        &  174 & 133 & Sloan~$i$ \\ 
&KeplerCam\tablenotemark{b} & 2010 Sep 06        &  129 & 163 & Sloan~$i$ \\
&FTN\tablenotemark{c}       & 2010 Oct 02        &  476 &  35 & Sloan~$i$ \\
\sidehead{\bf{\hatcurBBb}}
&HAT-5/089 & 2008 Oct--2009 Mar & 2779 & 330 & Sloan~$r$ \\
&HAT-8/089 & 2008 Oct--2009 Mar &   674 & 330 & Sloan~$r$ \\
&KeplerCam\tablenotemark{b} & 2011 Jan 01 & 199 & 73 & Sloan~$i$ \\
&KeplerCam\tablenotemark{b} & 2011 Jan 24 & 261 & 43 & Sloan~$i$ \\
[-1.5ex]
\enddata 
\tablenotetext{a}{
    \hatcur{} is in two HATNet fields, internally labeled as
    \hatcurfieldA{} and \hatcurfieldB. \hatcurBB{} is in one field, labeled \hatcurBBfield.
}
\tablenotetext{b}{
    Observations from the KeplerCam CCD camera on the \flwof{} telescope
}
\tablenotetext{c}{
    Observations from the 2.0\,m Faulkes Telescope North (FTN) at
    Haleakala Observatory in Hawaii
}
\ifthenelse{\boolean{emulateapj}}{
    \end{deluxetable*}
}{
    \end{deluxetable}
}
 
\reftabl{photobs} summarizes the photometric observations of \hatcurb{} and \hatcurBBb{}. Altogether 10,400 instrumental \band{R} 5.5 minute cadence photometric measurements of \hatcur{} (\hatcurCCgsc{}; \hatcurCCtwomass{}; $\alpha = \hatcurCCra$, $\delta = \hatcurCCdec$; J2000; V=\hatcurCCtassmv,\ \citealp{droege:2006}) were obtained with the HAT-6, HAT-7, HAT-8 and HAT-9 telescopes. 3,453 Sloan~$r$ measurements of \hatcurBB{} (\hatcurBBCCgsc{}; \hatcurBBCCtwomass{}; $\alpha = \hatcurBBCCra$, $\delta = \hatcurBBCCdec$; J2000; V=\hatcurBBCCtassmv) were obtained with the HAT-5 and HAT-8 telescopes. 

These observations revealed box-like transit signals in the light curves of the stars with an apparent depth of $\sim\hatcurLCdip$\,mmag and  $\sim\hatcurBBLCdip$\,mmag, and a period of $P=\hatcurLCPshort$\,days and $P=\hatcurBBLCPshort$\,days, respectively (see Figures \ref{fig:hatnet} and  \ref{fig:hatnetBB}).

\subsection{Reconnaissance Spectroscopy}
\label{sec:recspec}

High-resolution, low-S/N reconnaissance spectra were obtained for \hatcur{} and \hatcurBB{} using the Harvard-Smithsonian Center for Astrophysics (CfA) Digital Speedometer \citep[DS;][]{latham:1992} and the Tillinghast Reflector \'Echelle Spectrograph \citep[TRES;][]{furesz:2008} on the \flwos\ telescope as well as the FIber-fed \'Echelle Spectrograph (FIES) on the 2.5\,m Nordic Optical Telescope (NOT) at La Palma, Spain \citep{djupvik:2010}. The reconnaissance observations are listed in \reftabl{reconspecobs} and were used to estimate the effective temperature, surface gravity, and projected rotational velocity of the host stars as described by \cite{torres:2002}. The procedures used for reduction and extraction of high precision radial velocities of the FIES spectra are described in \cite{buchhave:2010}. 

The CfA DS reconnaissance observations revealed no detectable RV variation at the $\sim 1$\,\kms\ precision of the observations. The TRES and FIES reconnaissance observations showed radial velocities with a low RMS (\hatcurb\ showed an RMS of $12\ \ms$ and the \hatcurBB\ showed an RMS of $82\ \ms$), which both could be consistent with a planetary companion. Additionally the spectra are consistent with the host stars being single, slowly-rotating, dwarf stars.

\ifthenelse{\boolean{emulateapj}}{
    \begin{deluxetable*}{lllrrrr}
}{
    \begin{deluxetable}{lllrrrr}
}
\tablewidth{0pc}
\tabletypesize{\scriptsize}
\tablecaption{
    Summary of reconnaissance spectroscopy observations
    \label{tab:reconspecobs}
}
\tablehead{
    \multicolumn{2}{c}{Instrument}          &
    \multicolumn{1}{c}{Date}             &
    \multicolumn{1}{c}{$\teffstar$}         &
    \multicolumn{1}{c}{$\loggstar$}         &
    \multicolumn{1}{c}{$\vsini$}            &
    \multicolumn{1}{c}{$\gamma_{\rm RV}$\tablenotemark{a}} \\
    &
    &
    \multicolumn{1}{c}{(K)}                 &
    \multicolumn{1}{c}{(cgs)}               &
    \multicolumn{1}{c}{(\kms)}              &
    \multicolumn{1}{c}{(\kms)}
}
\startdata
\sidehead{\bf{\hatcurb}}
&DS   & 13 Jan 2009 & $5750$ & $4.5$ & $0$ & $+44.32$ \\
&DS   & 14 Jan 2009 & $5750$ & $4.5$ & $0$ & $+42.34$ \\
&FIES & 19 Jan 2009 & $5750$ & $4.5$ & $2$ & $+43.85$ \\
&FIES & 21 Jan 2009 & $5750$ & $4.5$ & $2$ & $+43.87$ \\
\sidehead{\bf{\hatcurBBb}}
&FIES & 17 Aug 2010 & $6000$ & $4.0$ & $6$ & $-21.67$ \\
&FIES & 17 Aug 2010 & $5750$ & $4.0$ & $6$ & $-21.76$ \\
&TRES & 21 Jan 2009 & $5750$ & $4.5$ & $6$ & $-21.59$ \\
[-1.5ex]
            
\enddata 
\tablenotetext{a}{
    The heliocentric RV of the target.
}
\ifthenelse{\boolean{emulateapj}}{
    \end{deluxetable*}
}{
    \end{deluxetable}
}

\subsection{Photometric follow-up observations}
\label{sec:phot}

\begin{figure}[!ht]
\plotone{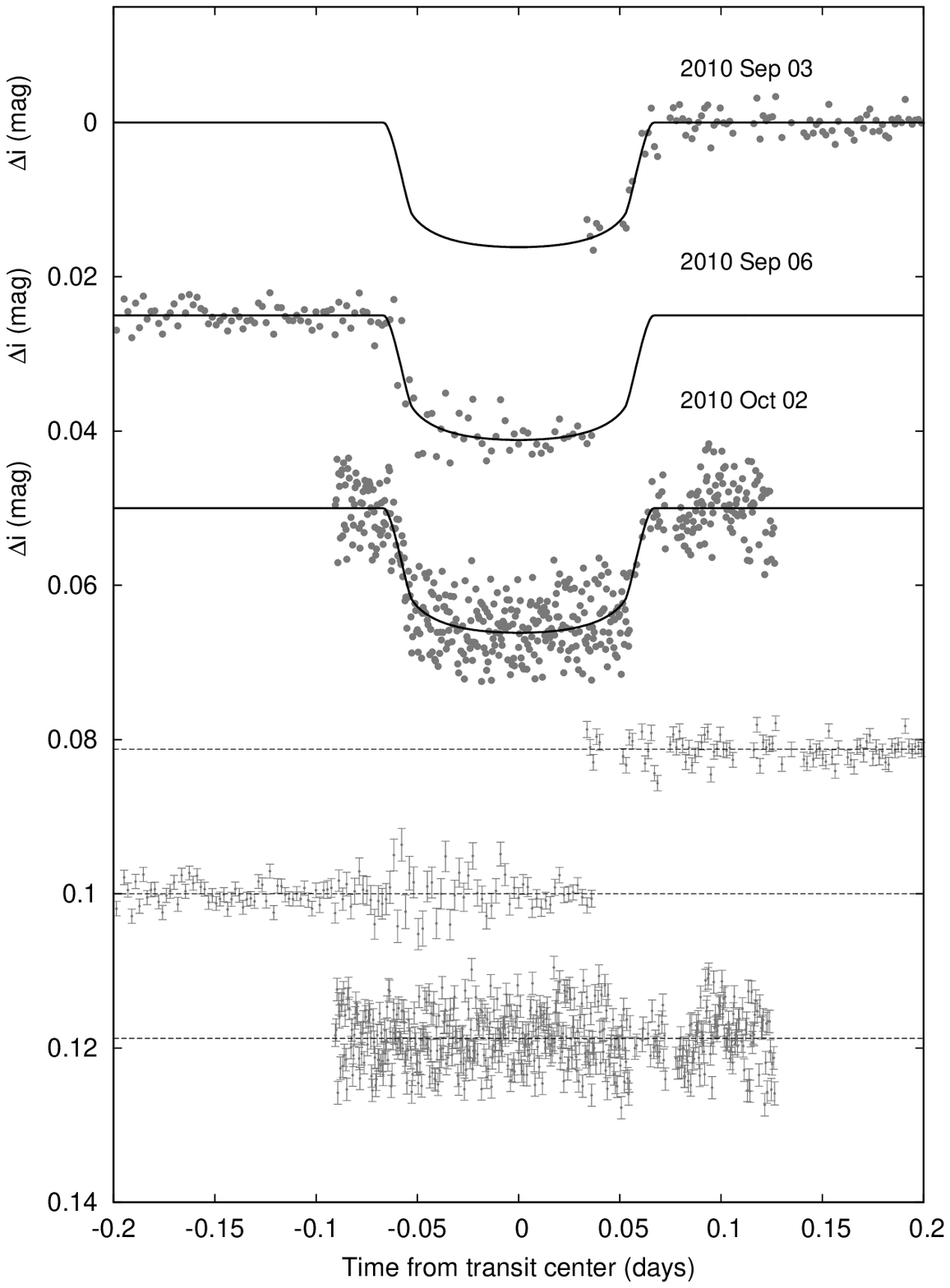}
\caption{
	Unbinned instrumental Sloan \band{i} transit \lcs{} of \hatcur{}, acquired
        with KeplerCam at the \flwof{} telescope on 2010 Sep 03 and
        2010 Sep 06, and with the Faulkes Telescope North on 2010 Oct
        02, from top to bottom.  The light curves have been cleaned of
        instrumental variations using the ELTG model (\refsec{globmod} and
        \citealt{bakos:2010}).
    Curves are
    displaced vertically for clarity.  Our best fit from the global
    modeling described in \refsecl{globmod} is shown by the solid
    lines.  Residuals from the fits are displayed at the bottom, in the
    same order as the top curves.  The error bars represent the photon
    and background shot noise, plus the readout noise.
\label{fig:lc}}
\end{figure}

\begin{figure}[!ht]
\plotone{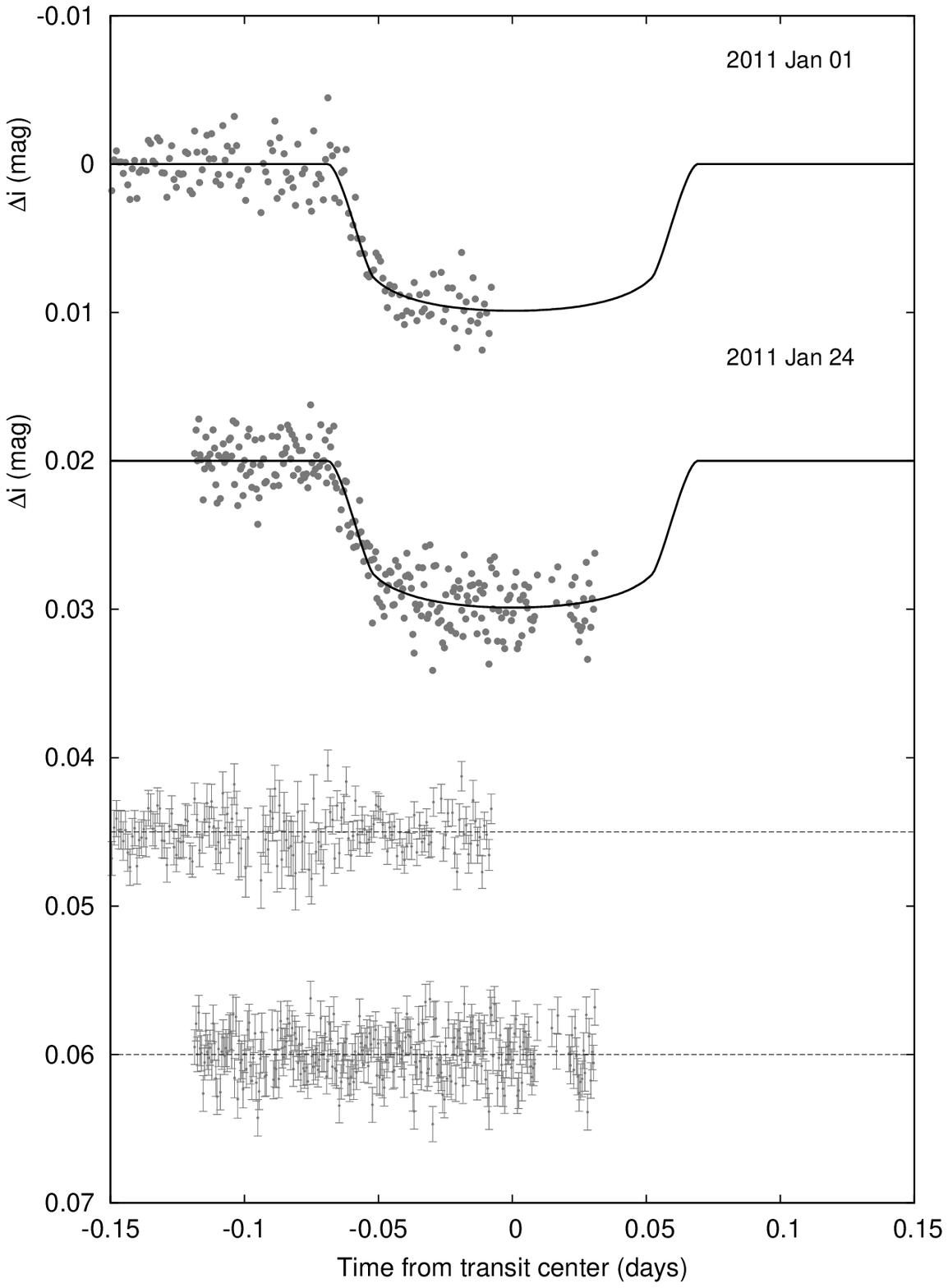}
\caption{
	Unbinned instrumental Sloan \band{i} transit \lcs{} of \hatcurBB{}, acquired
        with KeplerCam at the \flwof{} telescope on 2011 Jan 01 and
        Jan 24. See caption in Figure \ref{fig:lc} for details.
\label{fig:lcBB}}
\end{figure}
In-transit photometric observations of \hatcur{} and \hatcurBB{} were obtained with the KeplerCam CCD camera on the \flwof{} telescope and with the 2.0\,m Faulkes Telescope North (FTN) at Haleakala Observatory in Hawaii. These observations are summarized in \reftabl{photobs}.

The images were reduced to \lcs{} following the aperture photometry
procedure described by \citet{bakos:2010}. Figures \ref{fig:lc} and \ref{fig:lcBB} show the
final trend-filtered light curves together with the best-fit transit
model (see \refsecl{globmod}); the individual measurements are
provided in \reftabl{phfu} and \ref{tab:phfuBB}.

\begin{deluxetable}{lrrrr}
\tablewidth{0pc}
\tablecaption{High-precision differential photometry of 
	\hatcur\label{tab:phfu}}
\tablehead{
	\colhead{BJD} & 
	\colhead{Mag\tablenotemark{a}} & 
	\colhead{\ensuremath{\sigma_{\rm Mag}}} &
	\colhead{Mag(orig)\tablenotemark{b}} & 
	\colhead{Filter} \\
	\colhead{\hbox{~~~~(2,400,000$+$)~~~~}} & 
	\colhead{} & 
	\colhead{} &
	\colhead{} & 
	\colhead{}
}
\startdata
$ 55443.68956 $ & $   0.01257 $ & $   0.00104 $ & $  11.6842 $ & $ i$\\
$ 55443.69111 $ & $   0.01476 $ & $   0.00102 $ & $  11.6858 $ & $ i$\\
$ 55443.69266 $ & $   0.01656 $ & $   0.00102 $ & $  11.6878 $ & $ i$\\
$ 55443.69421 $ & $   0.01308 $ & $   0.00102 $ & $  11.6852 $ & $ i$\\
$ 55443.69576 $ & $   0.01364 $ & $   0.00102 $ & $  11.6776 $ & $ i$\\
$ 55443.70735 $ & $   0.01317 $ & $   0.00100 $ & $  11.6841 $ & $ i$\\
$ 55443.70889 $ & $   0.01369 $ & $   0.00100 $ & $  11.6852 $ & $ i$\\
$ 55443.71044 $ & $   0.00874 $ & $   0.00099 $ & $  11.6816 $ & $ i$\\
$ 55443.71197 $ & $   0.00762 $ & $   0.00100 $ & $  11.6786 $ & $ i$\\
$ 55443.71663 $ & $   0.00139 $ & $   0.00100 $ & $  11.6848 $ & $ i$\\
[-1.5ex]
\enddata
\tablenotetext{a}{
	The out-of-transit level has been subtracted. These magnitudes have
	been subjected to the External Parameter Decorrelation (EPD) technique and 
    Trend Filtering Algorithm (TFA) procedures \citep[see][]{bakos:2010}, 
	carried out
	simultaneously with the transit fit.
}
\tablenotetext{b}{
	Raw magnitude values without application of the EPD and TFA
	procedures.
}
\tablecomments{
    This table is available in a machine-readable form in the online
    journal.  A portion is shown here for guidance regarding its form
    and content.
}
\end{deluxetable}

\begin{deluxetable}{lrrrr}
\tablewidth{0pc}
\tablecaption{High-precision differential photometry of 
	\hatcurBB\label{tab:phfuBB}}
\tablehead{
	\colhead{BJD} & 
	\colhead{Mag\tablenotemark{a}} & 
	\colhead{\ensuremath{\sigma_{\rm Mag}}} &
	\colhead{Mag(orig)\tablenotemark{b}} & 
	\colhead{Filter} \\
	\colhead{\hbox{~~~~(2,400,000$+$)~~~~}} & 
	\colhead{} & 
	\colhead{} &
	\colhead{} & 
	\colhead{}
}
\startdata
$ 55563.67299 $ & $   0.00148 $ & $   0.00114 $ & $  10.8583 $ & $ i$\\
$ 55563.67366 $ & $   0.00031 $ & $   0.00115 $ & $  10.8581 $ & $ i$\\
$ 55563.67444 $ & $  -0.00043 $ & $   0.00099 $ & $  10.8565 $ & $ i$\\
$ 55563.67528 $ & $  -0.00158 $ & $   0.00099 $ & $  10.8556 $ & $ i$\\
$ 55563.67614 $ & $  -0.00072 $ & $   0.00097 $ & $  10.8564 $ & $ i$\\
$ 55563.67700 $ & $   0.00009 $ & $   0.00096 $ & $  10.8567 $ & $ i$\\
$ 55563.67784 $ & $  -0.00134 $ & $   0.00094 $ & $  10.8549 $ & $ i$\\
$ 55563.67870 $ & $   0.00242 $ & $   0.00095 $ & $  10.8600 $ & $ i$\\
$ 55563.67955 $ & $   0.00023 $ & $   0.00094 $ & $  10.8577 $ & $ i$\\
$ 55563.68042 $ & $   0.00100 $ & $   0.00095 $ & $  10.8581 $ & $ i$\\
[-1.5ex]
\enddata
\tablecomments{
   The notes of this table are identical to the notes in Table~\ref{tab:phfu}.
}
\end{deluxetable}

\subsection{High resolution, high S/N spectroscopy}
\label{sec:hispec}
We obtained high-resolution, high-S/N spectra of \hatcur{} using NOT/FIES
as well as HIRES \citep{vogt:1994} mounted on the Keck~I telescope on
Mauna Kea, Hawaii. A total of 10 high S/N FIES observations were
obtained between 3 and 18 October 2009, and a total of 8 HIRES
observations, including 6 through an iodine gas absorption cell and
two without the cell, were obtained between 1 October 2009 and 23
February 2010. For \hatcurBB{}, 9 HIRES spectra were obtained between 26 September and 14 December 2010, one of which was without the iodine cell.

\begin{figure} [ht]
\plotone{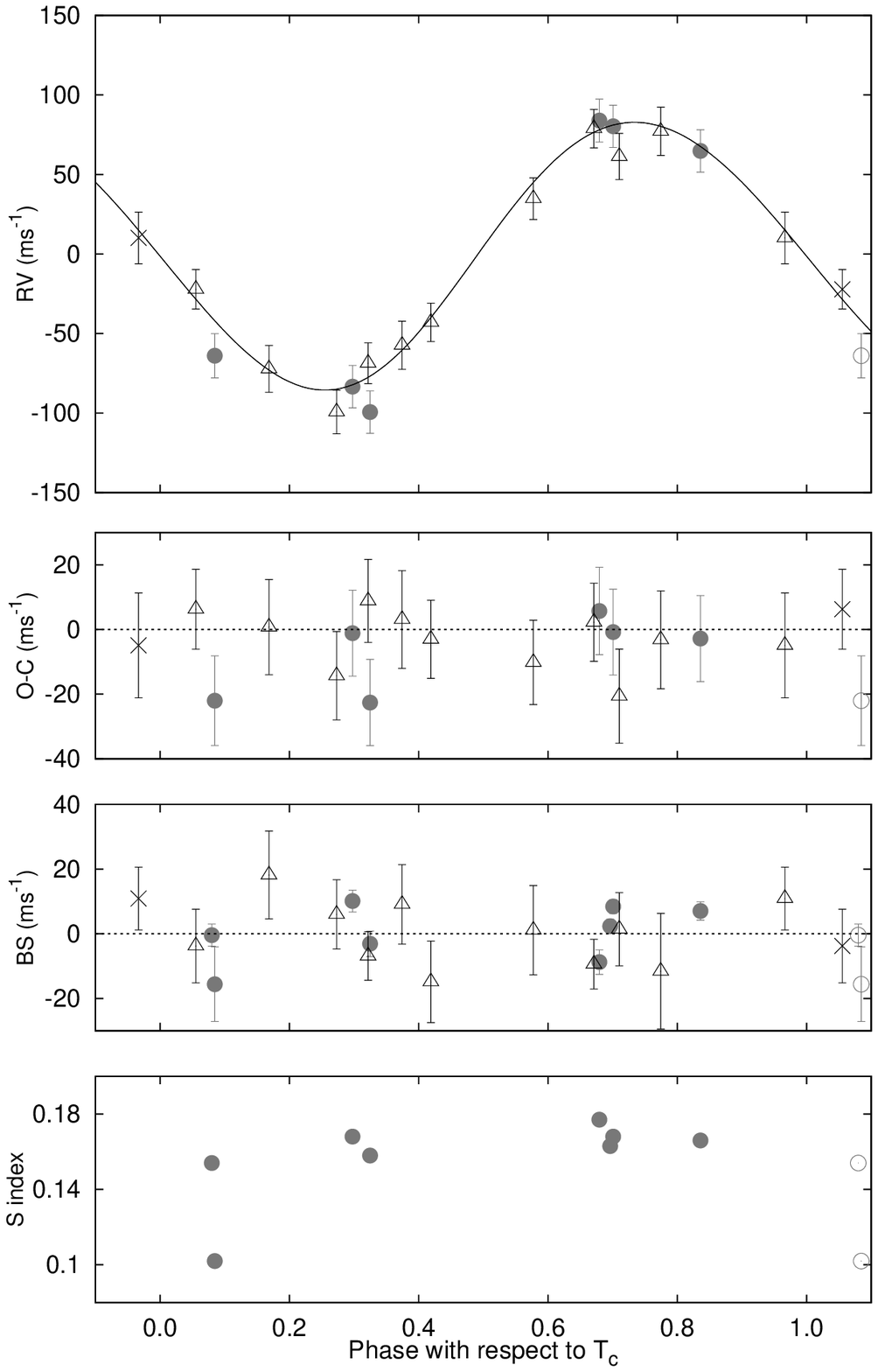}
\caption{
    {\em Top panel:} High-precision relative RV measurements for
    \hbox{\hatcur{}} from Keck/HIRES (filled circles) and
    NOT/FIES (open triangles) shown as a function of orbital
    phase, along with our best-fit model (see \reftabl{planetparam}). Repeated 
    points are
	shown as open circles and crosses for Keck/HIRES and NOT/FIES,
	respectively.
    Zero phase corresponds to the time of mid-transit.  The
    center-of-mass velocity has been subtracted. 
	{\em Second panel:} Velocity $O\!-\!C$ residuals from the best
        fit.  The error bars of the Keck/HIRES measurements include a
        component from astrophysical/instrumental jitter
        ($\hatcurRVjitterA$\,\ms; see \refsecl{globmod}); it was
        unnecessary to add jitter to the NOT/FIES measurements. The
        RMS of the residuals is $\hatcurRVfitrmsA$\,\ms\ and
        $\hatcurRVfitrmsB$\,\ms\ for the Keck/HIRES and NOT/FIES
        observations respectively.
	{\em Third panel:} Bisector spans (BS), with the mean value
    subtracted.  The measurement from the template spectrum is included.
	{\em Bottom panel:} Relative chromospheric activity index $S$
    measured from the Keck spectra.
    Note the different vertical scales of the panels.
\label{fig:rvbis}}
\end{figure}

\begin{figure} [ht]
\plotone{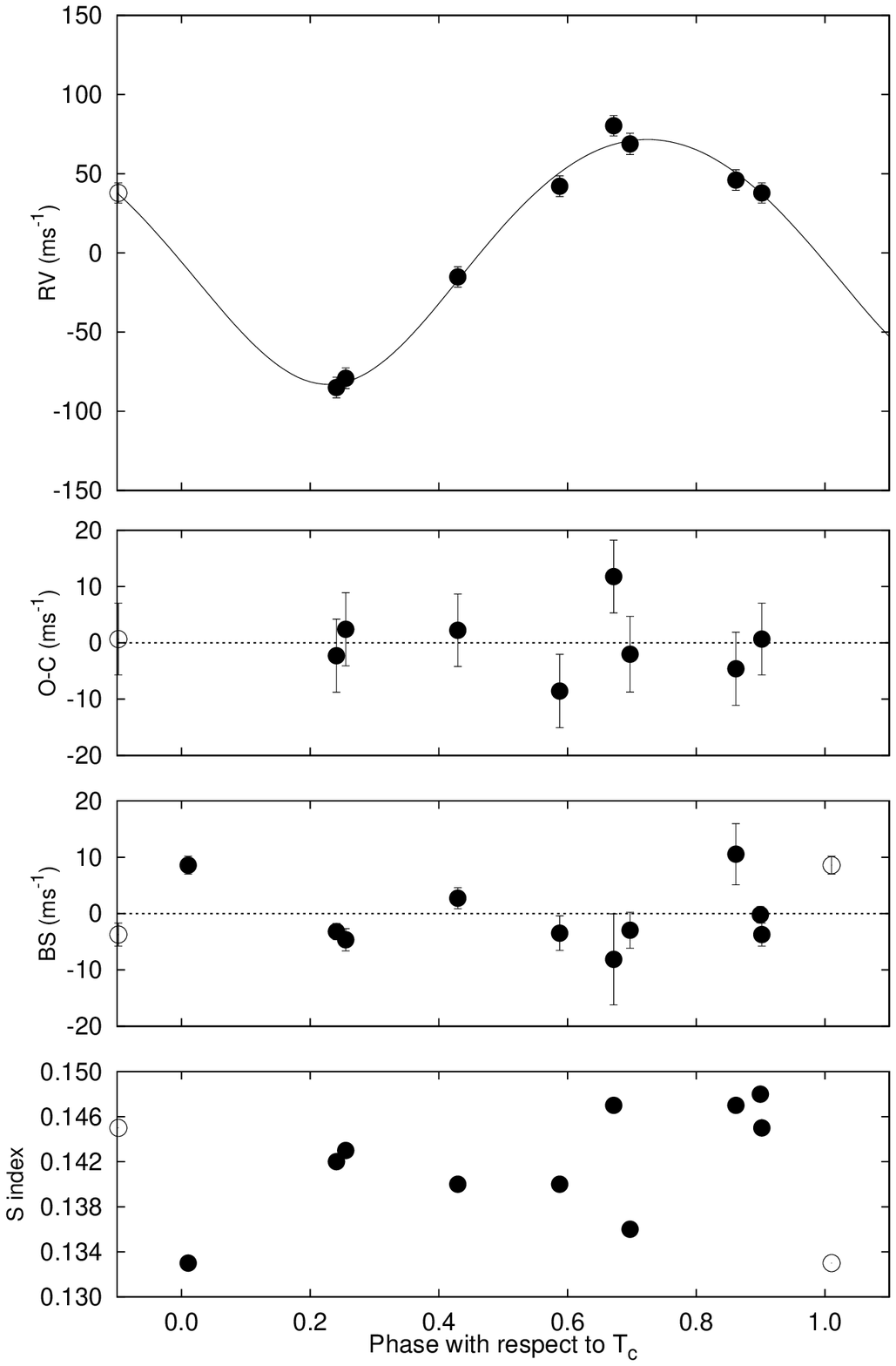}
\caption{
    {\em Top panel:} High-precision relative RV measurements for
    \hbox{\hatcurBB{}} from Keck/HIRES shown as a function of orbital
    phase, along with our best-fit model (see \reftabl{planetparam}).
    Zero phase corresponds to the time of mid-transit.  The
    center-of-mass velocity has been subtracted. Repeated points are
	shown as open circles.
	{\em Second panel:} Velocity $O\!-\!C$ residuals from the best
        fit.  The error bars of the Keck/HIRES measurements include a
        component from astrophysical/instrumental jitter
        ($\hatcurBBRVjitter$\,\ms; see \refsecl{globmod}). The
        RMS of the residuals is $\hatcurBBRVfitrms$\,\ms.
	{\em Third panel:} Bisector spans (BS), with the mean value
    subtracted.
	{\em Bottom panel:} Relative chromospheric activity index $S$
    measured from the Keck spectra.
    Note the different vertical scales of the panels.
\label{fig:rvbisBB}}
\end{figure}

For the FIES observations we used the medium-resolution fiber, giving
a resolution of $\lambda/\Delta\lambda\approx 46,\!000$ and a
wavelength coverage of $\sim$3600--7400\,\AA\@. The spectra were bracketed by ThAr calibration frames used to determine the fiducial wavelength calibration and were
reduced to relative RVs in the solar system barycentric frame
following a cross-correlation procedure described by
\cite{buchhave:2010}. For HIRES, we set the spectrometer slit to
$0\farcs86$, resulting in a resolving power of $\lambda/\Delta\lambda
\approx 55,\!000$ with a wavelength coverage of
$\sim$3800--8000\,\AA\@. We reduced these spectra to relative RVs in
the solar system barycentric frame following a procedure based on that
described by \cite{butler:1996}. The RV measurements and their
uncertainties are listed in \reftabl{rvs} and \ref{tab:rvsBB}. The period-folded data,
along with our best fit described below in \refsecl{analysis}, are
displayed in Figures \ref{fig:rvbis} and \ref{fig:rvbisBB}.

Following \cite{torres:2007} and \cite{queloz:2001}, we conducted a
bisector span analysis as described in \S 5 of \cite{bakos:2007a} of
the Keck and FIES spectra to rule out the possibility that the objects
are eclipsing binary star systems blended with light from a third star
rather than a planet. As seen in Figures \ref{fig:rvbis} and \ref{fig:rvbisBB} the bisector spans
show no variation in phase with the orbital period and have a total
scatter significantly less than the RV semi-amplitude, which supports the interpretation that the velocity variations are due to a planetary companions.

In the same figures, we also show the calibrated $S$ indices, which are a
measure of the chromospheric activity of the stars derived from the
flux in the cores of the \ion{Ca}{2} H and K lines
\citep{vaughan:1978,isaacson:2010}. We also measured the \rhk indices
for the systems to be $\logrhk=-4.984$ and $\logrhk=-5.105$, respectively, as described by
\citet{noyes:1984}. We do not detect any significant variation of the
indices correlated with orbital phases.

\ifthenelse{\boolean{emulateapj}}{
    \begin{deluxetable*}{lrrrrrrr}
}{
    \begin{deluxetable}{lrrrrrrr}
}
\tablewidth{0pc}
\tablecaption{
	Relative radial velocities, bisector spans, and activity index
	measurements of \hatcur{}. 
	\label{tab:rvs}
}
\tablehead{
	\colhead{BJD} & 
	\colhead{RV\tablenotemark{a}} & 
	\colhead{\ensuremath{\sigma_{\rm RV}}\tablenotemark{b}} & 
	\colhead{BS} & 
	\colhead{\ensuremath{\sigma_{\rm BS}}} & 
	\colhead{S\tablenotemark{c}} & 
        \colhead{Phase} &
        \colhead{Instrument}\\
	\colhead{\hbox{(2,400,000$+$)}} & 
	\colhead{(\ms)} & 
	\colhead{(\ms)} &
	\colhead{(\ms)} &
    \colhead{(\ms)} &
	\colhead{} &
        \colhead{} &
        \colhead{}
}
\startdata
$ 55107.11869 $ & $    83.86 $ & $     3.76 $ & $    -8.78 $ & $     3.78 $ & $    0.177 $ & $   0.680 $ & Keck \\
$ 55108.71101 $ & $   -72.31 $ & $    12.10 $ & $    18.20 $ & $    13.60 $ & \nodata      & $   0.168 $ & FIES \\
$ 55110.68494 $ & $    77.09 $ & $    12.60 $ & $   -11.60 $ & $    17.90 $ & \nodata      & $   0.774 $ & FIES \\
$ 55111.59971 $ & $   -22.21 $ & $     9.10 $ & $    -3.80 $ & $    11.40 $ & \nodata      & $   0.055 $ & FIES \\
$ 55112.63861 $ & $   -57.41 $ & $    12.60 $ & $     9.10 $ & $    12.30 $ & \nodata      & $   0.374 $ & FIES \\
$ 55113.60539 $ & $    78.79 $ & $     8.70 $ & $    -9.40 $ & $     7.70 $ & \nodata      & $   0.671 $ & FIES \\
$ 55114.56809 $ & $    10.09 $ & $    13.90 $ & $    10.90 $ & $     9.70 $ & \nodata      & $   0.967 $ & FIES \\
$ 55115.56626 $ & $   -99.31 $ & $    10.80 $ & $     6.00 $ & $    10.70 $ & \nodata      & $   0.273 $ & FIES \\
$ 55116.55720 $ & $    34.79 $ & $    10.00 $ & $     1.10 $ & $    13.80 $ & \nodata      & $   0.577 $ & FIES \\
$ 55122.55573 $ & $   -43.01 $ & $     8.70 $ & $   -14.90 $ & $    12.60 $ & \nodata      & $   0.419 $ & FIES \\
$ 55123.50498 $ & $    61.29 $ & $    11.90 $ & $     1.40 $ & $    11.30 $ & \nodata      & $   0.710 $ & FIES \\
$ 55125.49641 $ & $   -68.71 $ & $     9.70 $ & $    -6.90 $ & $     7.50 $ & \nodata      & $   0.322 $ & FIES \\
$ 55191.86077 $ & \nodata      & \nodata      & $     2.33 $ & $     2.14 $ & $    0.163 $ & $   0.696 $ & Keck \\
$ 55191.87545 $ & $    80.26 $ & $     2.74 $ & $     8.43 $ & $     2.12 $ & $    0.168 $ & $   0.701 $ & Keck \\
$ 55193.81948 $ & $   -83.44 $ & $     2.85 $ & $    10.12 $ & $     3.37 $ & $    0.168 $ & $   0.298 $ & Keck \\
$ 55198.83008 $ & $    64.79 $ & $     2.92 $ & $     7.03 $ & $     2.85 $ & $    0.166 $ & $   0.836 $ & Keck \\
$ 55229.73714 $ & $   -99.43 $ & $     3.16 $ & $    -3.11 $ & $     3.96 $ & $    0.158 $ & $   0.325 $ & Keck \\
$ 55251.74043 $ & \nodata      & \nodata      & $    -0.42 $ & $     3.40 $ & $    0.154 $ & $   0.080 $ & Keck \\
$ 55251.75531 $ & $   -64.03 $ & $     4.85 $ & $   -15.61 $ & $    11.55 $ & $    0.102 $ & $   0.085 $ & Keck \\
	[-1.5ex]
\enddata
\tablenotetext{a}{
        The zero-point of these velocities is arbitrary. An overall
        offset $\gamma_{\rm rel}$ fitted separately to the FIES and
        Keck velocities in \refsecl{globmod} has been subtracted.
}
\tablenotetext{b}{
	Internal errors excluding the component of astrophysical/instrumental jitter
    considered in \refsecl{globmod}.
}
\tablenotetext{c}{
	Relative chromospheric activity index, calibrated to the
	scale of \citet{vaughan:1978}.
}
\ifthenelse{\boolean{rvtablelong}}{
	\tablecomments{
		For the iodine-free template exposures there is no RV
		measurement, but the BS and S index can still be determined.
	}
}{
    \tablecomments{
		For the iodine-free template exposures there is no RV
		measurement, but the BS and S index can still be determined.
		This table is presented in its entirety in the
		electronic edition of the Astrophysical Journal.  A portion is
		shown here for guidance regarding its form and content.
	}
} 
\ifthenelse{\boolean{emulateapj}}{
    \end{deluxetable*}
}{
    \end{deluxetable}
}

\ifthenelse{\boolean{emulateapj}}{
    \begin{deluxetable*}{lrrrrrrr}
}{
    \begin{deluxetable}{lrrrrrrr}
}
\tablewidth{0pc}
\tablecaption{
	Relative radial velocities, bisector spans, and activity index
	measurements of \hatcurBB{}. 
	\label{tab:rvsBB}
}
\tablehead{
	\colhead{BJD} & 
	\colhead{RV\tablenotemark{a}} & 
	\colhead{\ensuremath{\sigma_{\rm RV}}\tablenotemark{b}} & 
	\colhead{BS} & 
	\colhead{\ensuremath{\sigma_{\rm BS}}} & 
	\colhead{S\tablenotemark{c}} & 
        \colhead{Phase} &
        \colhead{Instrument}\\
	\colhead{\hbox{(2,400,000$+$)}} & 
	\colhead{(\ms)} & 
	\colhead{(\ms)} &
	\colhead{(\ms)} &
    \colhead{(\ms)} &
	\colhead{} &
        \colhead{} &
        \colhead{}
}
\startdata
$ 55465.99028 $ &     \nodata  &   \nodata  & $-0.20$      & $1.44$      & $0.148 $       & $   0.900 $ & Keck \\
$ 55466.00284 $ & $    37.80 $ & $     2.16 $ & $-3.72$      & $2.05$      & $0.145 $       & $   0.902 $ & Keck \\
$ 55467.94293 $ & $   -85.06 $ & $     2.47 $ & $-3.19$     & $1.42$      & $0.142 $       & $   0.241 $ & Keck \\
$ 55469.02138 $ & $   -15.21 $ & $     2.31 $ & $2.73$      & $1.87$      & $0.140 $       & $   0.429 $ & Keck \\
$ 55469.92851 $ & $    42.10 $ & $     2.53 $ & $-3.47$      & $3.07$      & $0.140 $       & $   0.588 $ & Keck \\
$ 55490.91798 $ & $   -79.20 $ & $     2.51 $ & $    -9.19 $ & $     1.97 $ & $    0.143 $ & $   0.255 $ & Keck \\
$ 55500.96329 $ & \nodata      & \nodata      & $     3.48 $ & $     1.14 $ & $    0.133 $ & $   0.010 $ & Keck \\
$ 55521.91781 $ & $    80.29 $ & $     2.39 $ & $    -6.04 $ & $     7.56 $ & $    0.147 $ & $   0.672 $ & Keck \\
$ 55523.00463 $ & $    45.97 $ & $     2.49 $ & $    10.14 $ & $     5.01 $ & $    0.147 $ & $   0.862 $ & Keck \\
$ 55544.95590 $ & $    68.78 $ & $     3.04 $ & $    -7.18 $ & $     3.45 $ & $    0.136 $ & $   0.697 $ & Keck \\
	[-1.5ex]
\enddata
\tablenotetext{a}{
        The zero-point of these velocities is arbitrary. An overall
        offset $\gamma_{\rm rel}$ fitted separately to the FIES and
        Keck velocities in \refsecl{globmod} has been subtracted.
}
\tablenotetext{b}{
	Internal errors excluding the component of astrophysical/instrumental jitter
    considered in \refsecl{globmod}.
}
\tablenotetext{c}{
	Relative chromospheric activity index, calibrated to the
	scale of \citet{vaughan:1978}.
}
\ifthenelse{\boolean{rvtablelong}}{
	\tablecomments{
		For the iodine-free template exposures there is no RV
		measurement, but the BS and S index can still be determined.
	}
}{
    \tablecomments{
		For the iodine-free template exposures there is no RV
		measurement, but the BS and S index can still be determined.
		This table is presented in its entirety in the
		electronic edition of the Astrophysical Journal.  A portion is
		shown here for guidance regarding its form and content.
	}
} 
\ifthenelse{\boolean{emulateapj}}{
    \end{deluxetable*}
}{
    \end{deluxetable}
}

\section{Analysis}
\label{sec:analysis}

The analysis of the \hatcur{} and \hatcurBB{} systems, including determinations of the
properties of the host stars and planets, was carried out in a 
fashion similar to previous HATNet discoveries
\citep[e.g.][]{bakos:2010}. Below we briefly summarize the procedure
and results.

\subsection{Properties of the parent star}
\label{sec:stelparam}

Stellar atmospheric parameters were measured using our template
spectrum obtained with the Keck/HIRES instrument, and the analysis
package known as Spectroscopy Made Easy \citep[SME;][]{valenti:1996},
along with the atomic line database of \cite{valenti:2005}. SME
yielded {\em initial} values and uncertainties (which,
for the first two parameters, we have conservatively doubled to
include our estimates of the systematic errors based on prior experience \citep[see e.g.][]{noyes:2008}).
For \hatcur{} we found the following parameters: effective temperature $\teffstar=\hatcurSMEiteff$\,K, 
metallicity $\feh=\hatcurSMEizfeh$\,dex, projected rotational velocity $\vsini=\hatcurSMEivsin\,\kms$, and stellar surface gravity $\loggstar=\hatcurSMEilogg$\,(cgs). For \hatcurBB{} we found: effective temperature $\teffstar=\hatcurBBSMEiteff$\,K, metallicity $\feh=\hatcurBBSMEizfeh$\,dex, projected rotational velocity $\vsini=\hatcurBBSMEivsin\,\kms$, and stellar surface gravity $\loggstar=\hatcurBBSMEilogg$\,(cgs). 

Our initial values of \teffstar, \loggstar, and \feh\ were used to
determine the limb-darkening coefficients needed in the global
modeling of the follow-up photometry.  This modeling is briefly
summarized in \refsecl{globmod} with more details given by
\citet{bakos:2010}. Following \citet{sozzetti:2007} we used \arstar\
(the normalized semimajor axis), $\teffstar$ and $\feh$
together with the \hatcurisofull\ stellar evolution models
\citep{\hatcurisocite} to determine probability distributions of other
stellar properties (such as \loggstar, \rstar, and \mstar). Note that
\arstar\ is related to the mean stellar density, and is determined
from the analysis of the light curves and RV curves. This procedure
has been described in further detail by \cite{pal:2009b}.

\begin{figure}[!ht]
\plotone{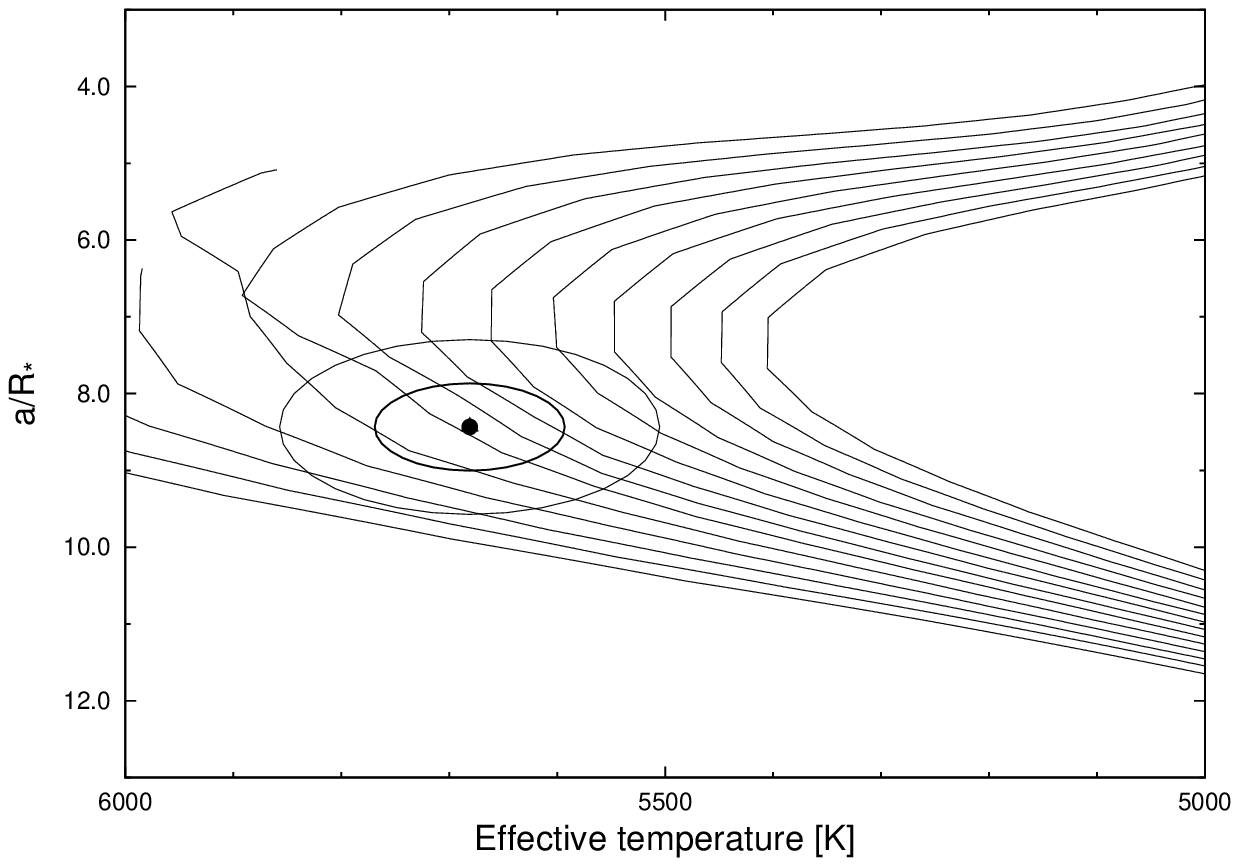}
\caption{
    Model isochrones from \cite{\hatcurisocite} for the measured
    metallicity of \hatcur, \feh = \hatcurSMEiizfehshort, and ages
    between 1 and 14\,Gyr in 1\,Gyr increments (left to right).  The
    adopted values of $\teffstar$ and \arstar\ are shown together with
    their 1$\sigma$ and 2$\sigma$ confidence ellipsoids.
\label{fig:iso}}
\end{figure}

\begin{figure}[!ht]
\plotone{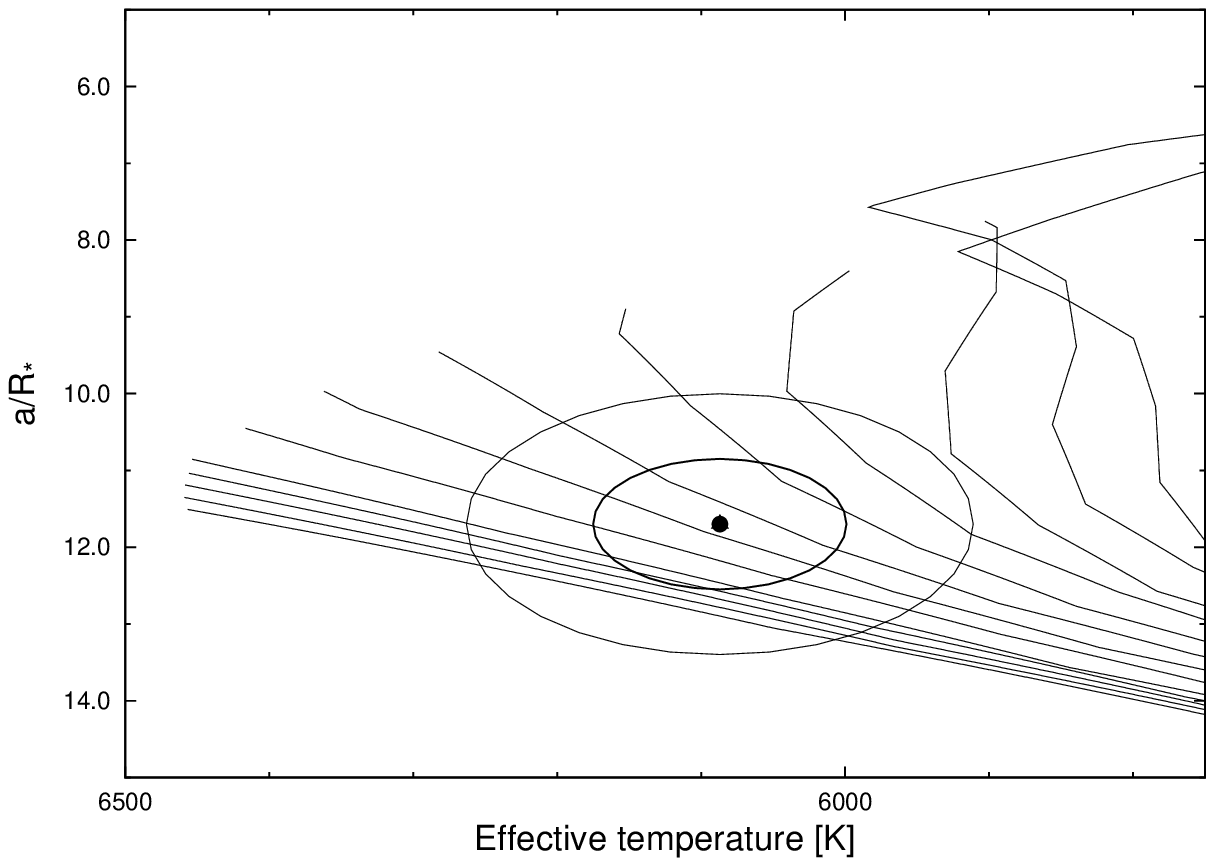}
\caption{
    Model isochrones from \cite{\hatcurisocite} for the measured
    metallicity of \hatcurBB, \feh = \hatcurBBSMEiizfehshort, and ages
    between 1 and 13\,Gyr in 1\,Gyr increments (left to right).  The
    adopted values of $\teffstar$ and \arstar\ are shown together with
    their 1$\sigma$ and 2$\sigma$ confidence ellipsoids.
\label{fig:isoBB}}
\end{figure}

The inferred location of the stars in a diagram of \arstar\ versus \teffstar\, together with \cite{\hatcurisocite} isochrones, is shown in Figures \ref{fig:iso} and \ref{fig:isoBB}. The SME results for the surface gravity yielded $\loggstar = \hatcurISOlogg$ for \hatcur{} and  $\loggstar = \hatcurBBISOlogg$ for \hatcurBB{}. Both values are consistent with our initial SME analysis and we therefore did not carry out a second iteration of SME and adopted these values for \loggstar\ and the values of \teffstar, \feh, and \vsini\ stated above as the final atmospheric properties of the star.  The stellar parameters are listed in \reftabl{stellar}. We find that \hatcur{} has a mass $\mstar\ = \hatcurISOmlong\,\msun$, radius $\rstar\ = \hatcurISOrlong\,\rsun$, and an age of \hatcurISOage\,Gyr and \hatcurBB{} has a mass \mstar\ = \hatcurBBISOmlong\,\msun, radius \rstar\ = \hatcurBBISOrlong\,\rsun, and an age of \hatcurBBISOage\,Gyr, according to these models.

As a check on the stellar evolution modeling we compare the observed
photometric color indices of the stars to the values predicted by the
modeling. We use the near-infrared magnitudes from the 2MASS Catalogue
\citep{skrutskie:2006}, which for \hatcur{} are
$J_{\rm 2MASS}=\hatcurCCtwomassJmag$, 
$H_{\rm 2MASS}=\hatcurCCtwomassHmag$ and 
$K_{\rm 2MASS}=\hatcurCCtwomassKmag$
and for \hatcurBB{} are
$J_{\rm 2MASS}=\hatcurBBCCtwomassJmag$, 
$H_{\rm 2MASS}=\hatcurBBCCtwomassHmag$ and 
$K_{\rm 2MASS}=\hatcurBBCCtwomassKmag$.
Converting these to the photometric system of the models (ESO) using
the transformations by \cite{carpenter:2001} we find a measured color
index of $J-K = \hatcurCCesoJKmag$ and $J-K = \hatcurBBCCesoJKmag$, respectively.  This is within 3$\sigma$ of the
value predicted from the isochrones of $J-K = \hatcurISOJK$ and $J-K = \hatcurBBISOJK$, respectively.

From the observed and predicted $J-K$ we find $E(J-K) = 0.077\pm0.036$ and $E(J-K) = 0.035\pm0.039$, respectively, and using \citet{cardelli:1989} we find $A(K) = 0.052\pm0.024$ and $A(K) = 0.024\pm0.026$, respectively, which gives a reddening corrected distance estimate, using the $K$ magnitude, of \hatcurXdist\,pc for \hatcur{} and \hatcurBBXdist\,pc for \hatcurBB{}. The uncertainty does not include difficult to quantify systematics in the model isochrones.

\begin{deluxetable*}{lrrl}
\tablewidth{0pc}
\tabletypesize{\scriptsize}
\tablecaption{
	Stellar parameters for \hatcur{} and \hatcurBB{}
	\label{tab:stellar}
}
\tablehead{
	\colhead{~~~~~~~~Parameter~~~~~~~~}	&
	\colhead{\hatcur{}} &
	\colhead{\hatcurBB{}} &
	\colhead{Source}
}
\startdata
\noalign{\vskip -3pt}
\sidehead{Spectroscopic properties}
~~~~$\teffstar$ (K)\dotfill         &  \hatcurSMEteff       &  \hatcurBBSMEteff       & SME\tablenotemark{a}\\
~~~~$\feh$\dotfill                  &  \hatcurSMEzfeh       &  \hatcurBBSMEzfeh       & SME                 \\
~~~~$\vsini$ (\kms)\dotfill         &  \hatcurSMEvsin       &  \hatcurBBSMEvsin       & SME                 \\
~~~~$\vmac$ (\kms)\dotfill          &  \hatcurSMEvmac       &  \hatcurBBSMEvmac       & SME                 \\
~~~~$\vmic$ (\kms)\dotfill          &  \hatcurSMEvmic       &  \hatcurBBSMEvmic       & SME                 \\
~~~~$\gamma_{\rm RV}$ (\kms)\dotfill&  \hatcurFIESgamma     &  \hatcurBBFIESgamma     & FIES                  \\
\sidehead{Photometric properties}
~~~~$V$ (mag)\dotfill               &  \hatcurCCtassmv      &  \hatcurBBCCtassmv      & TASS                \\
~~~~$V\!-\!I_C$ (mag)\dotfill       &  \hatcurCCtassvi      &  \hatcurBBCCtassvi      & TASS                \\
~~~~$J$ (mag)\dotfill               &  \hatcurCCtwomassJmag &  \hatcurBBCCtwomassJmag & 2MASS           \\
~~~~$H$ (mag)\dotfill               &  \hatcurCCtwomassHmag &  \hatcurBBCCtwomassHmag & 2MASS           \\
~~~~$K_s$ (mag)\dotfill             &  \hatcurCCtwomassKmag &  \hatcurBBCCtwomassKmag & 2MASS           \\
\sidehead{Derived properties}
~~~~$\mstar$ ($\msun$)\dotfill           &  \hatcurISOmlong      &  \hatcurBBISOmlong      & \hatcurisoshort+\hatcurlumind+SME \tablenotemark{b}\\
~~~~$\rstar$ ($\rsun$)\dotfill           &  \hatcurISOrlong      &  \hatcurBBISOrlong      & \hatcurisoshort+\hatcurlumind+SME         \\
~~~~$\loggstar$ (cgs)\dotfill            &  \hatcurISOlogg       &  \hatcurBBISOlogg       & \hatcurisoshort+\hatcurlumind+SME         \\
~~~~$\lstar$ ($\lsun$)\dotfill           &  \hatcurISOlum        &  \hatcurBBISOlum        & \hatcurisoshort+\hatcurlumind+SME         \\
~~~~$M_V$ (mag)\dotfill                  &  \hatcurISOmv         &  \hatcurBBISOmv         & \hatcurisoshort+\hatcurlumind+SME         \\
~~~~$M_K$ (mag,\hatcurjhkfilset)\dotfill &  \hatcurISOMK         &  \hatcurBBISOMK         & \hatcurisoshort+\hatcurlumind+SME         \\
~~~~Age (Gyr)\dotfill                    &  \hatcurISOage        &  \hatcurBBISOage        & \hatcurisoshort+\hatcurlumind+SME         \\
~~~~Distance (pc)\dotfill                &  \hatcurXdist         &  \hatcurBBXdist         & \hatcurisoshort+\hatcurlumind+SME\\
[-1.5ex]
\enddata
\tablenotetext{a}{
	SME = ``Spectroscopy Made Easy'' package for the analysis of
	high-resolution spectra \citep{valenti:1996}.  These parameters
	rely primarily on SME, but have a small dependence also on the
	iterative analysis incorporating the isochrone search and global
	modeling of the data, as described in the text.
}
\tablenotetext{b}{
	\hatcurisoshort+\hatcurlumind+SME = Based on the \hatcurisoshort\
    isochrones \citep{\hatcurisocite}, \hatcurlumind\ as a luminosity
    indicator, and the SME results.
}
\end{deluxetable*}

\subsection{Global modeling of the data}
\label{sec:globmod}

We modeled the HATNet photometry, the follow-up photometry, and the
high-precision RV measurements using the procedure described in detail
by \citet{bakos:2010,pal:2009}. To summarize, we simultaneously fit
these data using the \citet{mandel:2002} model for the follow-up \lcs\ from KeplerCam and FTN,
quadratic limb darkening coefficients interpolated from the tables by
\citet{claret:2004}, a simplified no-limb-darkening transit model for
the HATNet light curves, and a Keplerian orbit for the RVs. 

The free parameters in the fit include the time of the first and last transit center observed, the normalized planetary radius $p\equiv \rpl/\rstar$, the square of the impact parameter $b^2$, the reciprocal of the half duration of the
transit $\zrstar$, the RV semiamplitude $K$, Lagrangian elements $k\equiv e\cos\omega$, and $h \equiv e\sin\omega$, the HATNet blend factors (which account for possible dilution of the transit in the HATNet light curves from background stars due to the broad PSF ($\sim 26\arcsec$ FWHM) and the application of trend-filtering without signal reconstruction), the out-of-transit magnitudes for each HATNet field, the relative zero-points $\gamma_{\rm rel,HIRES}$ and $\gamma_{\rm rel,FIES}$ of the HIRES/Keck and FIES/NOT RVs, and 22 instrumental parameters from the ``ELTG'' model for systematics.

We inflated the errors for the HIRES velocities by adding in quadrature a
jitter of \hatcurRVjitterA\ \ms\ for \hatcurb{} and \hatcurBBRVjitter\ \ms\ for \hatcurBBb{} to the formal errors to achieve a $\chi^{2}/{\rm dof} = 1$. The internal uncertainty estimate of the FIES velocities, which is estimated by $\sigma=RMS(v)/\sqrt{N}$, where $v$ is the RVs of the individual orders and $N$ is the number of orders \citep[see][]{buchhave:2010}, might tend to slightly overestimate the uncertainties of the individual measurements and thus we do not need to add jitter to the FIES velocities to achieve a $\chi^{2}/{\rm dof} = 1$.

The derived planetary parameters can be found in Table \ref{tab:planetparam}. We find:
\begin{itemize}
\item \hatcurb{} -- The planet has a mass of $\mpl=\hatcurPPmlong\,\mjup$ and a radius of $\rpl=\hatcurPPrlong\,\rjup$, a mean density of $\rho_p=\hatcurPPrho$\,\gcmc.
\item \hatcurBBb{} -- The planet has a mass of $\mpl=\hatcurBBPPmlong\,\mjup$ and a radius of $\rpl=\hatcurBBPPrlong\,\rjup$, a mean density of $\rho_p=\hatcurBBPPrho$\,\gcmc.
\end{itemize}

We find an eccentricity of $e = \hatcurRVeccen$ and $e = \hatcurBBRVeccen$ for \hatcurb{} and \hatcurBBb{}, respectively, which are both consistent with a circular orbit.

\begin{deluxetable*}{lrr}
\tabletypesize{\scriptsize}
\tablecaption{Orbital and planetary parameters\label{tab:planetparam}}
\tablehead{
	\colhead{~~~~~~~~~~~~~~~Parameter~~~~~~~~~~~~~~~} &
	\colhead{\hatcurb} &
	\colhead{\hatcurBBb}
}
\startdata
\noalign{\vskip -3pt}
\sidehead{\Lc{} parameters}
~~~$P$ (days)             \dotfill    & $\hatcurLCP$          & $\hatcurBBLCP$              \\
~~~$T_c$ (${\rm BJD}$)                                                            
      \tablenotemark{a}   \dotfill    & $\hatcurLCT$          & $\hatcurBBLCT$              \\
~~~$T_{14}$ (days)                                                                
      \tablenotemark{a}   \dotfill    & $\hatcurLCdur$        & $\hatcurBBLCdur$            \\
~~~$T_{12} = T_{34}$ (days)                                                       
      \tablenotemark{a}   \dotfill    & $\hatcurLCingdur$     & $\hatcurBBLCingdur$         \\
~~~$\arstar$              \dotfill    & $\hatcurPPar$         & $\hatcurBBPPar$             \\
~~~$\zrstar$              \dotfill    & $\hatcurLCzeta$       & $\hatcurBBLCzeta$           \\
~~~$\rpl/\rstar$          \dotfill    & $\hatcurLCrprstar$    & $\hatcurBBLCrprstar$        \\
~~~$b^2$                  \dotfill    & $\hatcurLCbsq$        & $\hatcurBBLCbsq$            \\
~~~$b \equiv a \cos i/\rstar$                                                     
                          \dotfill    & $\hatcurLCimp$        & $\hatcurBBLCimp$            \\
~~~$i$ (deg)              \dotfill    & $\hatcurPPi$          & $\hatcurBBPPi$              \\

\sidehead{Limb-darkening coefficients \tablenotemark{b}}
~~~$a_i$ (linear term)    \dotfill    & $\hatcurLBii$         & $\hatcurBBLBii$             \\
~~~$b_i$ (quadratic term) \dotfill    & $\hatcurLBiii$        & $\hatcurBBLBiii$            \\
                                                                                
\sidehead{RV parameters}                                                        
~~~$K$ (\ms)              \dotfill    & $\hatcurRVK$          & $\hatcurBBRVK$              \\
~~~$e \rm{cos}\omega$\tablenotemark{c}                                                
                          \dotfill    & $\hatcurRVk$          & $\hatcurBBRVk$              \\
~~~$e \rm{sin}\omega$\tablenotemark{c}                                                
                          \dotfill    & $\hatcurRVh$          & $\hatcurBBRVh$              \\
~~~$e$                    \dotfill    & $\hatcurRVeccen$      & $\hatcurBBRVeccen$          \\
~~~$\omega$ (deg)         \dotfill    & $\hatcurRVomega$      & $\hatcurBBRVomega$          \\
~~~RV jitter HIRES (\ms)\tablenotemark{d}                 
                          \dotfill    & $\hatcurRVjitterA$    & $\hatcurBBRVjitter $        \\
                                                          
\sidehead{Secondary eclipse parameters}                   
~~~$T_s$ (BJD)            \dotfill    & $\hatcurXsecondary$  & $\hatcurBBXsecondary$       \\
~~~$T_{s,14}$             \dotfill    & $\hatcurXsecdur$      & $\hatcurBBXsecdur$           \\
~~~$T_{s,12}$             \dotfill    & $\hatcurXsecingdur$  & $\hatcurBBXsecingdur$       \\
                                                                                    
\sidehead{Planetary parameters}                                                     
~~~$\mpl$ ($\mjup$)       \dotfill    & $\hatcurPPmlong$      & $\hatcurBBPPmlong$           \\
~~~$\rpl$ ($\rjup$)       \dotfill    & $\hatcurPPrlong$      & $\hatcurBBPPrlong$           \\
~~~$C(\mpl,\rpl)$                                                                   
    \tablenotemark{e}     \dotfill    & $\hatcurPPmrcorr$     & $\hatcurBBPPmrcorr$         \\
~~~$\rhopl$ (\gcmc)       \dotfill    & $\hatcurPPrho$        & $\hatcurBBPPrho$            \\
~~~$\log g_p$ (cgs)       \dotfill    & $\hatcurPPlogg$       & $\hatcurBBPPlogg$           \\
~~~$a$ (AU)               \dotfill    & $\hatcurPParel$       & $\hatcurBBPParel$           \\
~~~$T_{\rm eq}$ (K)       \dotfill    & $\hatcurPPteff$       & $\hatcurBBPPteff$           \\
~~~$\Theta$\tablenotemark{f}\dotfill  & $\hatcurPPtheta$      & $\hatcurBBPPtheta$            \\
~~~$F_{per}$ ($10^{\hatcurPPfluxperidim}$\ergscmsq) \tablenotemark{g}
                          \dotfill    & $\hatcurPPfluxperi$   & $\hatcurBBPPfluxperi$         \\
~~~$F_{ap}$  ($10^{\hatcurPPfluxapdim}$\ergscmsq) \tablenotemark{g} 
                          \dotfill    & $\hatcurPPfluxap$     & $\hatcurBBPPfluxap$           \\
~~~$\langle F \rangle$ ($10^{\hatcurPPfluxavgdim}$\ergscmsq) 
\tablenotemark{g}         \dotfill    & $\hatcurPPfluxavg$    & $\hatcurBBPPfluxavg$          \\
[-1.5ex]
\enddata
\tablenotetext{a}{
    \ensuremath{T_c}: Reference epoch of mid transit that minimizes the
    correlation with the orbital period. BJD is calculated from UTC.
	\ensuremath{T_{14}}: total transit duration, time between first to
	last contact;
	\ensuremath{T_{12}=T_{34}}: ingress/egress time, time between first
	and second, or third and fourth contact.
}
\tablenotetext{b}{
	Values for a quadratic law, adopted from the tabulations by
    \cite{claret:2004} according to the spectroscopic (SME) parameters
    listed in \reftabl{stellar}.
}
\tablenotetext{c}{
    Lagrangian orbital parameters derived from the global modeling, and
    primarily determined by the RV data.
}
\tablenotetext{d}{
    This jitter was added to the Keck/HIRES measurements only. The
    NOT/FIES observations are consistent with no jitter.  
}
\tablenotetext{e}{
	Correlation coefficient between the planetary mass \mpl\ and radius
	\rpl.
}
\tablenotetext{f}{
	The Safronov number is given by $\Theta = \frac{1}{2}(V_{\rm
	esc}/V_{\rm orb})^2 = (a/\rpl)(\mpl / \mstar )$
	\citep[see][]{hansen:2007}.
}
\tablenotetext{g}{
	Incoming flux per unit surface area, averaged over the orbit.
}
\end{deluxetable*}



\section{Discussion}
\label{sec:discussion}
We present the discovery of the transiting planets \hatcurb\ and \hatcurBBb. Figure \ref{fig:exomr} shows the two planets on a mass--radius diagram comparing them to the other known TEPs. We discuss the properties of each planet below.

\subsection{\hatcurb}
\hatcurb\ has a period of $P=\hatcurLCP$\,d, a mass of $\mpl=\hatcurPPm\,\mjup$ and a radius of $\rpl=\hatcurPPr\,\rjup$, leading to a mean density $\rho_p=\hatcurPPrho$\,\gcmc.  We find an eccentricity of $e = \hatcurRVeccen$, which is consistent with a circular orbit. Both Kepler--6b \citep{dunham:2010} with $\mpl=0.67\ \mjup$, $\rpl=1.32\ \rjup$, $P=3.23\ \rm{d}$ \citep{kipping:2010} and Kepler--8b \citep{jenkins:2010} with $\mpl=0.60\ \mjup$, $\rpl=1.42\ \rjup$, $P=3.52\ \rm{d}$ \citep{kipping:2010} are quite similar to \hatcurb\ in terms of their mass and periods, but have slightly larger radii. A significant number of other transiting planets have very comparable properties to \hatcurb. In fact, in histograms of mass and radius of the discovered transiting planets with a bin--size of $0.5\ \mjup$ and $0.25\ \rjup$, respectively, \hatcurb\ resides in the most populated bins.

\begin{figure*}[!ht]
\plotone{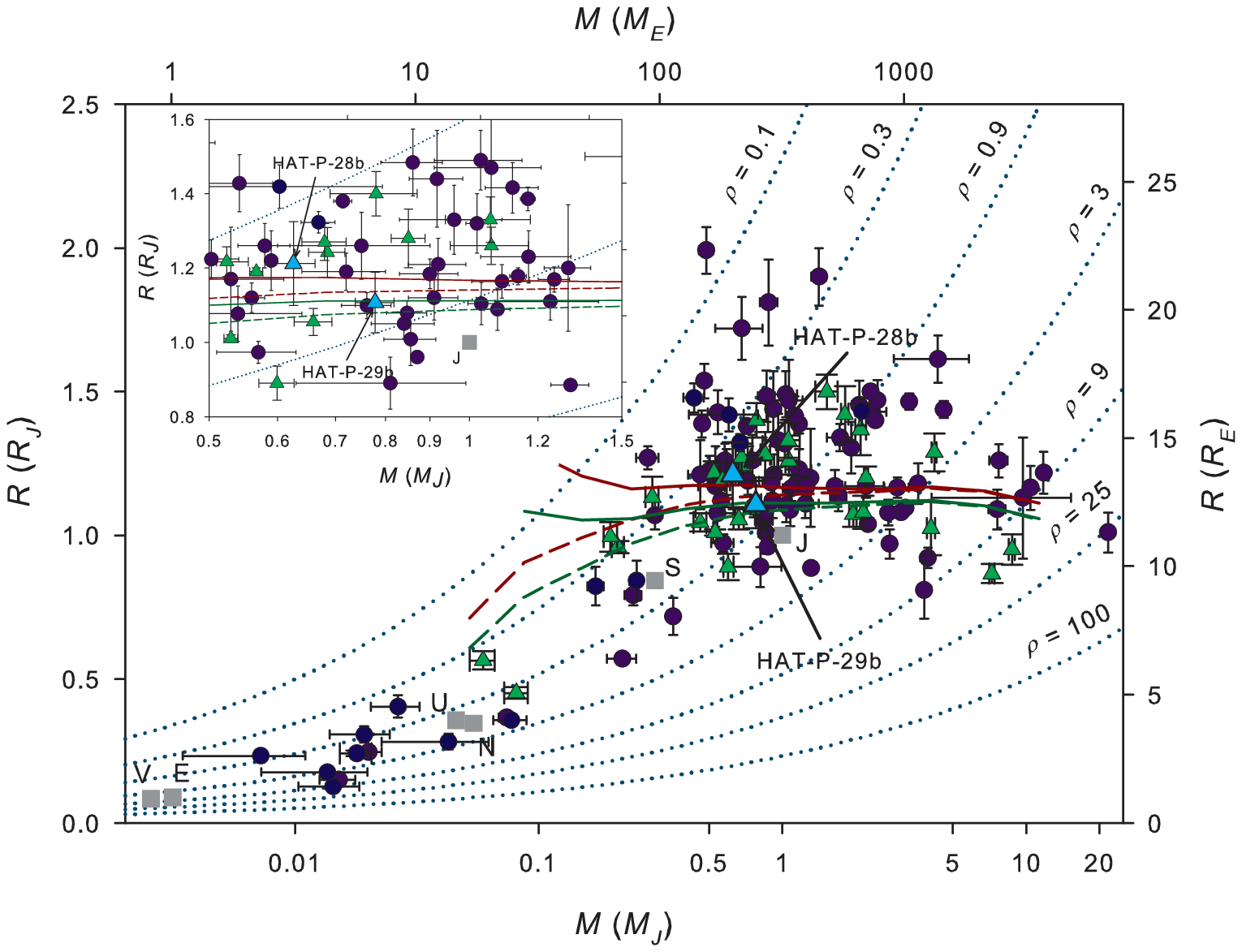}
\caption{
   Mass--radius diagram  of currently  known TEPs. HATNet planets are show as 
   green triangles and TEPs from other surveys  are shown as blue circles. 
   \hatcurb\ and \hatcurBBb\ are shown as large light blue triangles. The Solar system planets 
   are shown as  filled gray squares.  The insert in the top left corner is a zoom of the region where  \hatcurb\ and \hatcurBBb\ reside. Isodensity curves (in  $\rm{g\ cm^{-3}}$) 
   are plotted as dotted lines. Overlaid are planetary 1.0 Gyr (brown, the upper 
   set of  lines) and  4.5 Gyr (green, the  lower set of lines)  isochrones from 
   \citet{fortney:2007} for H/He-dominated planets with core mases of  $M_{c}=0\ 
   \mearth$ (solid) and $M_{c}=10\ \mearth$ (dashed), interpolated to the  solar 
   equivalent semimajor axis  of \hatcurb. Given the estimated age of 
   \hatcurISOage\ Gyr, \hatcurb\ may be slightly larger than what is allowed by 
   the theoretical models. \hatcurBBb\ is consistent with a H/He dominated planet with negligible core mass but also with a 1 Gyr model with a core mass of $M_{c}=10\ \mearth$. 
\label{fig:exomr}}
\end{figure*}

We have compared \hatcurb\ to the theoretical models from \citet{fortney:2007} 
by interpolating the models to the solar equivalent semimajor axis of $a = 
\hatcurPParel\ \rm{AU}$, the result of which can be seen overplotted in Figure 
\ref{fig:exomr}. Given the estimated age of \hatcurISOage\ Gyr, we find that 
\hatcurb\ may be slightly larger than what is allowed by the theoretical models, 
but otherwise consistent with models of a H/He-dominated planet with a 
negligible core mass.

\subsection{\hatcurBBb}
\hatcurBBb\ has a period of $P=\hatcurBBLCP$\,d, a mass of $\mpl=\hatcurBBPPm\,\mjup$ and a radius of $\rpl=\hatcurBBPPr\,\rjup$, leading to a mean density $\rho_p=\hatcurBBPPrho$\,\gcmc, and has an eccentricity of $e = \hatcurBBRVeccen$, which is consistent with a circular orbit. Again, \hatcurBBb\ is quite similar to a number of other transiting planets and is almost identical in mass and radius to TrES-1 \citep[$\mpl=0.75\ \mjup$, $\rpl=1.08\ \rjup$, $P=3.03\ \rm{d}$, ][]{alonso:2004}, albeit with a slightly longer period.

When compared with the \citet{fortney:2007} planetary models, \hatcurBBb\ could be consistent with a 4.5 Gyr model for a H/He-dominated planet with a negligible core mass but also with a 1 Gyr model with a core mass of $M_{c}=10\ \mearth$.

\acknowledgements 
HATNet operations have been funded by NASA grants NNG04GN74G, NNX08AF23G and SAO 
IR\&D grants.  Work of G.\'A.B. was supported by the Postdoctoral Fellowship of 
the NSF Astronomy and Astrophysics Program (AST-0702843 and AST-0702821, 
respectively).  GT acknowledges partial support from NASA grant NNX09AF59G.  We 
acknowledge partial support also from the Kepler Mission under NASA Cooperative 
Agreement NCC2-1390 (D.W.L., PI).  G.K.~thanks the Hungarian Scientific Research 
Foundation (OTKA) for support through grant K-81373.  This research has made use 
of Keck telescope time granted through NOAO (program A201Hr) and NASA (N018Hr, 
N167Hr). This paper uses observations obtained with facilities of the Las 
Cumbres Observatory Global Telescope and the Nordic Optical Telescope, operated 
on the island of La Palma jointly by Denmark, Finland, Iceland, Norway, and 
Sweden, in the Spanish Observatorio del Roque de los Muchachos of the Instituto 
de Astrofisica de Canarias. Thank you to Johan Fynbo for help on the distance 
estimate.





\begin{thebibliography}{}


\bibitem[Alonso et al.(2004)]{alonso:2004} Alonso, R., Brown, T.~M., 
Torres, G., Latham, D.~W., Sozzetti, A., Mandushev, G., Belmonte, J.~A., 
Charbonneau, D., Deeg, H.~J., Dunham, E.~W., O'Donovan, F.~T., 
\& Stefanik, R.~P.~2004, \apjl, 613, L153 




\bibitem[Bakos et al.(2004)]{bakos:2004}
 Bakos, G.~\'A., Noyes, R.~W., Kov\'acs, G., Stanek, K.~Z.,
 Sasselov, D.~D., \& Domsa, I.~2004, \pasp, 116, 266

\bibitem[Bakos et al.(2007)]{bakos:2007a}
 Bakos, G.~\'A., et al.~2007, \apj, 670, 826

\bibitem[Bakos et al.(2010)]{bakos:2010} Bakos, G.~{\'A}., et al.~2010,
\apj, 710, 1724

of the first four months of data 
\bibitem[Borucki et al.(2011)]{borucki:2011} Borucki, W.~J., Koch, D.~G., 
Basri, G., Batalha, N., Brown, T.~M., Bryson, S.~T., Caldwell, D., 
Christensen-Dalsgaard, J., Cochran, W.~D., DeVore, E., Dunham, E.~W., 
Gautier, T.~N., III, Geary, J.~C., Gilliland, R., Gould, A., Howell, S.~B., 
Jenkins, J.~M., Latham, D.~W., Lissauer, J.~J., Marcy, G.~W., Rowe, J., 
Sasselov, D., Boss, A., Charbonneau, D., Ciardi, D., Doyle, L., Dupree, 
A.~K., Ford, E.~B., Fortney, J., Holman, M.~J., Seager, S., Steffen, J.~H., 
Tarter, J., Welsh, W.~F., Allen, C., Buchhave, L.~A., Christiansen, J.~L., 
Clarke, B.~D., D{\'e}sert, J.-M., Endl, M., Fabrycky, D., Fressin, F., 
Haas, M., Horch, E., Howard, A., Isaacson, H., Kjeldsen, H., Kolodziejczak, 
J., Kulesa, C., Li, J., Machalek, P., McCarthy, D., MacQueen, P., Meibom, 
S., Miquel, T., Prsa, A., Quinn, S.~N., Quintana, E.~V., Ragozzine, D., 
Sherry, W., Shporer, A., Tenenbaum, P., Torres, G., Twicken, J.~D., Van 
Cleve, J., \& Walkowicz, L.~2011, arXiv:1102.0541 

\bibitem[Buchhave et al.(2010)]{buchhave:2010} Buchhave, L.~A. et al.~2010, \apj, 720, 1118 

\bibitem[Butler et al.(1996)]{butler:1996} 
Butler, R.~P.~et al.~1996, \pasp, 108, 500

\bibitem[Cardelli et al.(1989)]{cardelli:1989} Cardelli, J.~A., Clayton, G.~C., \& Mathis, J.~S.~1989, \apj, 345, 245 


\bibitem[Carpenter(2001)]{carpenter:2001} Carpenter, J.~M.~2001, \aj, 121, 2851 

\bibitem[Charbonneau et al.(2005)]{charbonneau:2005} Charbonneau, D., et al.~2005, \apj, 626, 523 

\bibitem[Claret(2004)]{claret:2004}
 Claret, A.~2004, \aap, 428, 1001

\bibitem[Djupvik \& Andersen(2010)]{djupvik:2010} Djupvik, A.~A., \&
  Andersen, J.\ 2010, in ``Highlights of Spanish Astrophysics V''
  eds. J.~M.~Diego, L.~J.~Goicoechea, J.~I.~Gonz\'alez-Serrano, \&
  J.~Gorgas (Springer: Berlin), p. 211

\bibitem[Droege et al.(2006)]{droege:2006}
Droege, T.~F., Richmond, M.~W., \& Sallman, M.~2006, \pasp, 118, 1666

\bibitem[Dunham et al.(2010)]{dunham:2010} Dunham, E.~W. et al.~2010, \apjl, 713, L136 

\bibitem[Fortney et al.(2007)]{fortney:2007} Fortney, J.~J., Marley, M.~S., \& Barnes, J.~W.~2007, \apj, 659, 1661

\bibitem[F{\H u}r{\'e}sz (2008)]{furesz:2008} F{\H u}r{\'e}sz,~G. 2008 Ph.D. thesis, University of Szeged, Hungary


\bibitem[Hansen \& Barman(2007)]{hansen:2007} Hansen, B.~M.~S., \& Barman, T.~2007, \apj, 671, 861 

\bibitem[Isaacson \& Fischer(2010)]{isaacson:2010} Isaacson, H., \& Fischer, D.~2010, \apj, 725, 875 

\bibitem[Jenkins et al.(2010)]{jenkins:2010} Jenkins, J.~M et al.~2010, \apj, 724, 1108 

\bibitem[Kipping \& Bakos(2010)]{kipping:2010} Kipping, D.~M., \& Bakos, G.~{\'A}.~2010, arXiv:1004.3538 

\bibitem[Latham(1992)]{latham:1992}
 Latham, D.~W.~1992, in IAU Coll.~135, Complementary Approaches to
 Double and Multiple Star Research, ASP Conf.~Ser.~32, 
 eds.~H.~A.~McAlister \& W.~I.~Hartkopf (San Francisco: ASP), 110

\bibitem[Mandel \& Agol(2002)]{mandel:2002}
 Mandel, K., \& Agol, E.~2002, \apjl, 580, L171

\bibitem[Morton 
\& Johnson(2011)]{morton:2011} Morton, T.~D., \& Johnson, J.~A.~2011, \apj, 729, 138 

\bibitem[Noyes et al.(1984)]{noyes:1984} Noyes, R.~W., Hartmann, L.~W.,
Baliunas, S.~L., Duncan, D.~K., \& Vaughan, A.~H.~1984, \apj, 279, 763

\bibitem[Noyes et al.(2008)]{noyes:2008} Noyes, R.~W., Bakos, G.~{\'A}., 
Torres, G., P{\'a}l, A., Kov{\'a}cs, G., Latham, D.~W., Fern{\'a}ndez, 
J.~M., Fischer, D.~A., Butler, R.~P., Marcy, G.~W., Sip{\H o}cz, B., 
Esquerdo, G.~A., Kov{\'a}cs, G., Sasselov, D.~D., Sato, B., Stefanik, R., 
Holman, M., L{\'a}z{\'a}r, J., Papp, I., 
\& S{\'a}ri, P.~2008, \apjl, 673, L79 




\bibitem[P\'al(2009)]{pal:2009}
P\'al, A.\ 2009, \mnras, 396, 1737

\bibitem[P\'al(2009b)]{pal:2009b}
P\'al, A.\ 2009b, arXiv:0906.3486, PhD thesis, E\"{o}tv\"{o}s Lor\'{a}nd University

\bibitem[Queloz et al.(2001)]{queloz:2001} Queloz, D., Henry, G.~W., 
Sivan, J.~P., Baliunas, S.~L., Beuzit, J.~L., Donahue, R.~A., Mayor, M., 
Naef, D., Perrier, C., \& Udry, S.~2001, \aap, 379, 279 

\bibitem[Skrutskie et al.(2006)]{skrutskie:2006} Skrutskie, M.~F., et 
al.~2006, \aj, 131, 1163

\bibitem[Sozzetti et al.(2007)]{sozzetti:2007}
 Sozzetti, A.~et al.~2007, \apj, 664, 1190


insights into the dynamical origins of hot Jupiters 
\bibitem[Triaud et al.(2010)]{triaud:2010} Triaud, A.~H.~M.~J., Collier 
Cameron, A., Queloz, D., Anderson, D.~R., Gillon, M., Hebb, L., Hellier, 
C., Loeillet, B., Maxted, P.~F.~L., Mayor, M., Pepe, F., Pollacco, D., 
S{\'e}gransan, D., Smalley, B., Udry, S., West, R.~G., 
\& Wheatley, P.~J.~2010, \aap, 524, A25 

\bibitem[Torres et al.(2002)]{torres:2002}
Torres, G., Neu\"hauser, R., \& Guenther, E.\ W. 2002, \aj, 123, 1701

\bibitem[Torres et al.(2007)]{torres:2007}
 Torres, G.~et al.~2007, \apjl, 666, 121

\bibitem[Valenti \& Piskunov(1996)]{valenti:1996}
 Valenti, J.~A., \& Piskunov, N.~1996, \aaps, 118, 595

\bibitem[Valenti \& Fischer(2005)]{valenti:2005}
 Valenti, J.~A., \& Fischer, D.~A. 2005, \apjs, 159, 141

\bibitem[Vaughan, Preston \& Wilson(1978)]{vaughan:1978}
Vaughan, A.~H., Preston, G.~W., \& Wilson, O.~C.~1978, \pasp, 90, 267

\bibitem[Vogt et al.(1994)]{vogt:1994}
 Vogt, S.~S.~et al.~1994, Proc.~SPIE, 2198, 362

\bibitem[Yi et al.(2001)]{yi:2001}
 Yi, S.~K.~et al.~2001, \apjs, 136, 417

\bibitem[Winn et al.(2005)]{winn:2005} Winn, J.~N., Noyes, R.~W., Holman, 
M.~J., Charbonneau, D., Ohta, Y., Taruya, A., Suto, Y., Narita, N., Turner, 
E.~L., Johnson, J.~A., Marcy, G.~W., Butler, R.~P., 
\& Vogt, S.~S.~2005, \apj, 631, 1215 

\end{thebibliography}
\end{document}